# Effect of various electron and hole transport layers on the performance of CsPbI$_3$-based perovskite solar cells: A numerical investigation in DFT, SCAPS-1D, and wxAMPS frameworks

M. Khalid Hossain[1, 2*], Mirza Humaun Kabir Rubel[3], G.F. Ishraque Toki[4], Intekhab Alam[5], Md. Ferdous Rahman[6], H. Bencherif[7]

[1]Institute of Electronics, Atomic Energy Research Establishment, Bangladesh Atomic Energy Commission, Dhaka 1349, Bangladesh
[2]Department of Advanced Energy Engineering Science, Interdisciplinary Graduate School of Engineering Science, Kyushu University, Fukuoka 816-8580, Japan
[3]Department of Materials Science and Engineering, University of Rajshahi, Rajshahi 6205, Bangladesh
[4]College of Materials Science and Engineering, Donghua University, Shanghai 201620, China
[5]Department of Mechanical and Manufacturing Engineering, University of Calgary, Calgary, AB T2N 1N4, Canada
[6]Department of Electrical and Electronic Engineering, Begum Rokeya University, Rangpur 5400, Bangladesh
[7]HNS-RE2SD, Higher National School of Renewable Energies, Environment and Sustainable Development, Batna 05078, Algeria

Correspondence: * khalid.baec@gmail.com; khalid@kyudai.jp; ORCID: 0000-0003-4595-6367

**Abstract**

CsPbI$_3$ has recently received tremendous attention as a possible absorber of perovskite solar cells (PSCs). However, CsPbI$_3$-based PSCs have yet to achieve the high performance of the hybrid PSCs. In this work, we performed a density functional theory (DFT) study using the Cambridge Serial Total Energy Package (CASTEP) code for the cubic CsPbI$_3$ absorber to compare and evaluate its structural, electronic, and optical properties. The calculated electronic band gap ($E_g$) using the GGA-PBE approach of CASTEP was 1.483 eV for this CsPbI$_3$ absorber. Moreover, the computed density of states (DOS) exhibited the dominant contribution from the Pb-5$d$ orbital, and most charge also accumulated for the Pb atom as seen from the electronic charge density map. Fermi surface calculation showed multiband character, and optical properties were computed to investigate the optical response of CsPbI$_3$. Furthermore, we used IGZO, SnO$_2$, WS$_2$, CeO$_2$, PCBM, TiO$_2$, ZnO, and C$_{60}$ as the electron transport layers (ETLs), and Cu$_2$O, CuSCN, CuSbS$_2$, Spiro-MeOTAD, V$_2$O$_5$, CBTS, CFTS, P3HT, PEDOT: PSS, NiO, CuO, and CuI as the hole transport layers (HTLs) to identify the best HTL/CsPbI$_3$/ETL combinations using the SCAPS-1D solar cell simulation software. Among 96 device structures, the best-optimized device structure, ITO/TiO$_2$/CsPbI$_3$/CBTS/Au was identified, which exhibited an efficiency of 17.9%. The effect of absorber and ETL thickness, series resistance, shunt resistance, and operating temperature was also evaluated for the six best devices along with their corresponding generation rate, recombination rate, capacitance-voltage, current density-voltage, and quantum efficiency characteristics. The obtained results from SCAPS-1D were also compared with wxAMPS simulation software.

# Contents

# Graphical Abstract

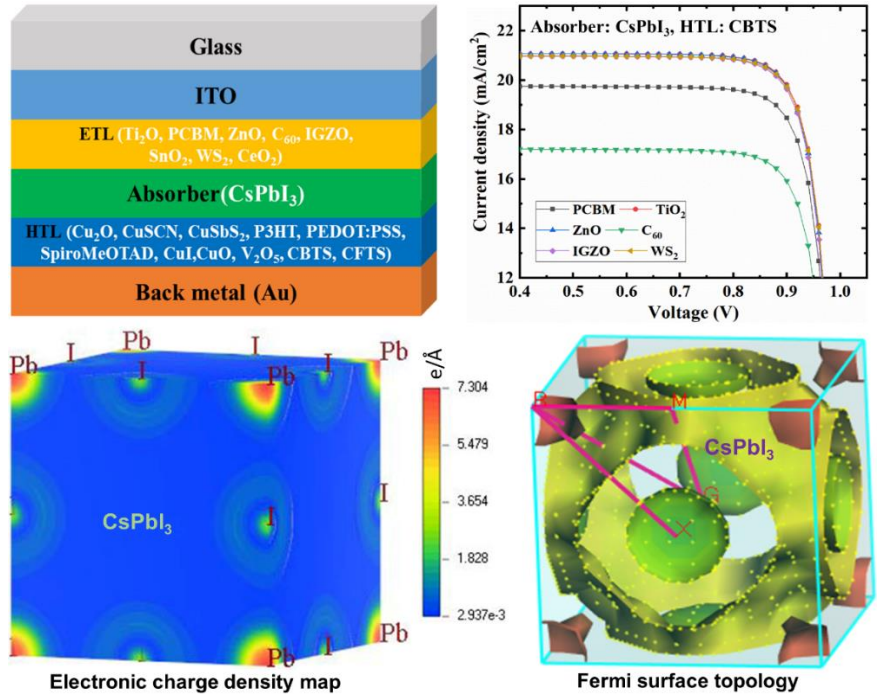

### List of abbreviations

| | | | |
|---|---|---|---|
| $\mu_h$ | Hole mobility | $N_t$ | Defect density |
| $\mu_n$ | Electron mobility | $N_V$ | VB effective density of states |
| CASTEP | Cambridge serial total energy package | PBE | Perdew-burke-ernzerhif |
| CBTS | Copper barium thiostannate | PCE | Power conversion efficiency |
| DFT | Density functional theory | PDOS | Partial density of states |
| DOS | Density of states | PSC | Perovskite solar cell |
| EC | Conduction band energy | PV | Photovoltaic |
| $E_g$ | Band gap | $R(\omega)$ | Reflectivity |
| ETL | Electron transport layer | $R_S$ | Series resistance |
| EV | Valence band energy | $R_{SH}$ | Shunt resistance |
| FF | Fill factor | SCAPS-1D | One dimensional solar cell capacitance simulator |
| GGA | Generalized gradient approximation | TCO | Transparent conductive oxide |
| HTL | Hole transport layer | TDOS | Total density of states |
| ITO | Indium doped tin oxide | $V_{OC}$ | Open voltage current |
| $J_e$ | Electron flux | wxAMPS | Widget provided analysis of microelectronic and photonic structures |
| $J_h$ | Hole flux | X | Electron affinity |
| $J_{SC}$ | Short circuit current density | XC | Exchange-correlation functional |
| J-V | Current density-voltage | α | Absorption coefficient |
| k | Extinction coefficient | $\alpha(\omega)$ | Optical absorption |
| L | Loss function | $\delta(\omega)$ | Conductivity |
| $L(\omega)$ | Electron energy loss function | $\varepsilon(\omega)$ | Dielectric function |
| n | Refractive index | $\varepsilon_1$ | Real |
| $n(\omega)$ | Refractive index | $\varepsilon_2$ | Imaginary |
| $N_A$ | Shallow uniform acceptor density | $\varepsilon_r$ | Dielectric permittivity |
| $N_C$ | CB effective density of states | λ | Light wavelength |
| $N_D$ | Shallow uniform donor density | σ | Electrical conductivity |
| $\omega_p$ | Plasma frequency | $V_{bi}$ | Built-in potential |
| MS | Mott-Schottky | $\Delta E_f$ | Formation energy |
| C-V | Capacitance-voltage | $E_F$ | Fermi level |
| QE | Quantum efficiency | | |

# 1 Introduction

The 21$^{st}$ century is experiencing a massive demand for power, with a substantial amount of need being met by fossil fuels, which pollute the environment and cause the greenhouse effect [1–3]. A lot of research is being carried out to study the alternate, ecologically suitable energy-generating options, with the solar cell being the most popular and cost-effective option [4–10]. To address future industrial needs, current photovoltaic (PV) investigations are focusing on developing highly efficient devices at cheap manufacturing costs. Even though various studies focus on exceeding the current greatest practicable performance, the PV characteristics are still far below the maximum value [5,11–14]. Energy shortages and pollutions are the major barriers to the creation of a sustainable society [4]. Therefore, the PV solar cell can be employed as a viable photoelectric conversion system, making this the most potential solution to the problems.

Hybrid halide PSCs, the third generation of PV cells, have received a lot of attention because of their straightforward production technique and cheaper cost [1]. The composition of halide perovskite absorber material is denoted as $ABX_3$, where A is a monovalent cation, B is a divalent metal, and X is a halide anion [9]. Limited exciton binding energy [2], prolonged carrier lifetime, and high diffusion length [1] along with an excellent absorption coefficient [10] are the important characteristics of these hybrid perovskites. Because of these outstanding PV characteristics, the power conversion efficiency (PCE) of these PSCs has enhanced from 3.8% to 25.5% [15,16] from the year 2009 to 2020. But because of their poor stability, they are intrinsically unsuited for long-term application in the outdoors [3,17,18].

Due to the similar photoelectric characteristics and improved thermal stability compared to the hybrid perovskites, inorganic $CsPbI_3$-based PSCs have recently gained a lot of attention [19–21]. With a bandgap of around 1.7 eV and exceptional thermal stability, $CsPbI_3$ is regarded as an excellent material for creating inorganic PSCs [21]. Initially, $CsPbI_3$-based PSCs had poor PCE compared to the hybrid PSCs. Since then, $CsPbI_3$-based PSCs have obtained a PCE reaching up to 19.03% [22]. The optoelectronic characteristics of $CsPbI_3$ are hindered by the quick transformation of the black perovskite to the yellow orthorhombic non-perovskites at ambient temperature and humidity [23]. In a recent simulation conducted with one dimensional solar cell capacitance simulator (SCAPS-1D) software, $CsPbI_3$-based PSC attained a PCE of 15.6% [24], where ZnO and $CuSbS_2$ were employed as the ETL and HTL, respectively. As shown by the large disparity between the practical and theoretical results, further studies are required to find a better-matched ETL and HTL for $CsPbI_3$-based PSC.

For the reliability and performance of the PSCs, both ETL and HTL have crucial roles [24,25]. Inorganic and organic materials such as $Al_2O_3$, PCBM, $C_{60}$, $ZrO_2$, $TiO_2$, $SnO_2$, $WO_x$, ZnO, $I_2O_3$, $BaTiO_3$, and $PbTiO_3$ nanoparticles have been used as the ETL in the PSCs over the years [24–29]. Due to the wide band gap, better energy level, and environmental stability, $TiO_2$ is one of the highly utilized ETLs [30]. Also, the suitable band gap, good stability, and desirable band bending make $TiO_2$ as one of the best ETLs to be used with the $CsPbI_3$ perovskite absorber material. The $TiO_2$ nanostructures, including nanotubes, nanosheets, nanoparticles, nanorods, and nanowires, have been employed as the ETL in high-performing PSCs [25,26]. The $TiO_2$ ETL has been widely used to remove and transfer the photo-generated electrons and protect the transparent conductive oxide (TCO) from interacting with the HTL [31]. As a result, there is an imbalance between the electron flux ($J_e$) and hole flux ($J_h$) owing to the lower charge mobility of $TiO_2$ as compared to the charge mobilities of the most commonly employed HTLs [32]. Therefore, one of the primary issues with the $TiO_2$-based planar PSCs is the hysteresis of the current density-voltage ($J$-$V$) relationship [33].

High-performance PV devices are difficult to fabricate because of the poor efficiency of carrier extraction [34]. The performance, durability, and production cost of the solar cells are influenced by the HTL and its neighboring surfaces [25,35]. Small molecule HTLs enable the solar cells to perform better, but they don't have enough thermal and photo-stability. Contrarily, polymeric HTLs are attractive because they are stable at high temperatures, repel water, and are compatible with other materials [25,26]. But these HTLs may have disadvantages such as poor optoelectronic properties, which can cause less efficiency. The HTLs made of inorganic materials are more chemically stable and cheaper than those fabricated from organic materials [36]. The rising need for low-cost solar cell production encourages the use of earth-abundant, air-stable thin-film materials such as copper barium thiostannate (CBTS) as the HTL, instead of the more commonly used $Cu_2O$, CuSCN, $CuSbS_2$, NiO, P3HT, PEDOT: PSS, Spiro-MeOTAD, CuI, CuO, and $V_2O_5$ HTLs, due to its tunable bandgap and good light-absorbing capabilities. Among the cation elements, the noncentrosymmetric crystal structure and substantial atomic size variations of CBTS ensure favorable characteristics to improve the PCE of the PV cells [35,37,38]. Previous studies have also mentioned that the CBTS HTL has good properties because of its appropriate absorption coefficient and electron affinity [35,38].

In this work, initially, the symmetric, electronic, as well as optical properties of the $CsPbI_3$ absorber have been evaluated by the first-principle computations in the framework of DFT by employing CASTEP software. Then we attempted to determine the optimized combinations of ETL and HTL with the $CsPbI_3$ absorber to enhance the device's performance by performing SCAPS-1D numerical investigations as it is time-consuming and expensive to investigate all possible combinations experimentally. However, very few theoretical investigations have been conducted for the $CsPbI_3$ absorber as compared to the other absorbers, and most of them are limited to enhancing the device performance by modifying various parameters only. Therefore, we have conducted a study on several HTLs ($Cu_2O$, CuSCN, $CuSbS_2$, NiO, P3HT, PEDOT: PSS, Spiro-MeOTAD, CuI, CuO, $V_2O_5$, CBTS, and CFTS) and ETLs (PCBM, $TiO_2$, ZnO, $C_{60}$, IGZO, $SnO_2$, $WS_2$, and $CeO_2$) (**Table 1** and **2**) to find the best combination. After that, for the best six devices, we checked the effect of absorber and ETL thickness, series resistance, shunt resistance, and working temperature on the device performance along with their band diagram, capacitance-voltage (*C-V*), generation rate, recombination rate, *J-V*, and quantum efficiency (*QE*) characteristics. Finally, the performance parameters of these structures have been validated with the widget provided analysis of microelectronic and photonic structures (wxAMPS) simulation software.

## 2 Computational study and numerical simulations

### 2.1 First principal calculations of CsPbI3 absorber using DFT

Herein, the computational studies using DFT in the framework of the CASTEP code [39] were accomplished to compare and correlate the properties. In this approach, the generalized gradient approximation (GGA) to the exchange-correlation potential with the parameter Perdew-Burke-Ernzerhif (PBE) [39,40] was applied, where ultrasoft pseudopotential rituality of the Vanderbilt type [41] was set for all elements interactions between valence electrons and ion ores. As the selection of exchange-correlation functionals (XCs) is an important parameter for DFT calculations, thus we optimized the structure with $Pm\bar{3}m$ symmetry of cubic phase. Broyden-Fletcher-Goldfarb-Shannon (BFGS) algorithm [42] was exploited with different XCs to find the minimum energy state of the complete stable structure. The formation energy ($\Delta E_f$, eV/atom) in the optimized structure and the calculated lattice constants are mainly compared with the reported experimental data. The best-produced data by XC was applied to compute all the characteristics of $CsPbI_3$ material. Cutoff energy of 520 eV was set in the wave function for the computational study of the $CsPbI_3$ solar absorber. Monkhorst-Pack grid [43] of $12 \times 12 \times 12$ (*k*-point) was applied for modeling the irreducible Brillouin zone. On the other hand, a greater scale of *k*-point mesh $17 \times 17 \times 17$ was employed to observe the electronic charge density map and Fermi surface topology, evidently. In these calculations, $1 \times 10^{-6}$ eV/atom (full energy) was used for the convergence tolerances of geometry optimization, 0.03 eV/Å was the highest potency on atoms, 0.001 Å was the maximal displacement of atoms, and 0.05 GPa was the maximum stress.

### 2.2 SCAPS-1D numerical simulation

The simulation model makes it easier to comprehend the fundamentals of solar cells and identifies the major factors that influence how well they function. The SCAPS-1D software [44,45] numerically solves the one-dimensional equations that influence the conduction of the semiconductor materials' charge carriers when they are in a stable state. Poisson's equation, which describes the connection between the electric field (E) of a p-n junction and the space charge density (ρ), is presented in **Eq. (1)**:

$$\frac{d}{dx}\left(-\varepsilon(x)\frac{d\psi}{dx}\right) = q[p(x) - n(x) + N_d^+(x) - N_a^-(x)] \quad (1)$$

Here, *ε* denotes the permittivity, *q* denotes the electron charge, *ψ* denotes the electrostatic potential, *n* is the total electron density, *p* represents the total hole density, $N_d^+$ represents the ionized donor-like doping concentration, and $N_a^-$ denotes the ionized acceptor-like doping concentration. **Eqs. (2)** and **(3)**, regarded as the electron and holes continuity equations, are given below:

$$\frac{\partial j_n}{\partial x} = q(R_n - G + \frac{\partial n}{\partial t}) \quad (2)$$

$$\frac{\partial j_p}{\partial x} = -q(R_p - G + \frac{\partial p}{\partial t}) \quad (3)$$

Here, $j_n$ symbolizes the electron density, $j_p$ stands for the hole's current density, $R_n$ denotes the net recombination rates for the electrons per unit volume, $R_p$ is the net recombination rate for the holes per unit volume, and $G$ is the generation rate per unit volume.

In this work, the new $E_g$-sqrt model, which is the updated version of the previous SCAPS model (the conventional sqrt $(h\vartheta - E_g)$ law model), was used to calculate the absorption constant for each layer. This model can be found from the "Tauc laws". The $E_g$-sqrt model is described in **Eq. (4)**:

$$\alpha(h\upsilon) = (\alpha_0 + \beta_0 \frac{E_g}{h\upsilon})\sqrt{\frac{h\upsilon}{E_g} - 1} \quad (4)$$

Here, $\alpha$ denotes the optical absorption constant, $h\upsilon$ implies the photon energy, and $E_g$ symbolizes the bandgap. The model constants, $\alpha_0$ and $\beta_0$, have the dimension of the absorption constant (e.g., cm$^{-1}$) and are related to the conventional model constants A and B by the following **Eqs. (5)** and **(6)**:

$$\alpha_0 = A\sqrt{E_g} \quad (5)$$

$$\beta_0 = \frac{B}{\sqrt{E_g}} \quad (6)$$

### 2.3 wxAMPS numerical simulation

In wxAMPS numerical simulation, Poisson's equation in 1-D space is used, which is given in **Eq. (7)** [46]:

$$\frac{d}{dx}\left(-\varepsilon(x)\frac{d\psi'}{dx}\right) = q.[p(x) - n(x) + N_D^+(x) - N_A^-(x) + pt(x) - nt(x)] \quad (7)$$

Where, the electrostatic potential is denoted by $\psi'$, the concentrations of the ionized donor-like and acceptor-like dopants are denoted by $N_D^+$ and $N_A^-$, respectively, the free electron is indicated by *n*, the free hole is symbolized by *p*, the trapped electron is represented by nt, and the trapped hole is denoted by pt. All of them are the variables of the position coordinate, x. The free electron carriers in the delocalized states of the conduction bands are represented by the following formula (**Eq. (8)**) [47]:

$$\frac{1}{q}\left(\frac{dJn}{dx}\right) = -G_{op}(x) + Rx \quad (8)$$

The continuity equation in the case of the free holes in the delocalized states of the valence band has been illustrated in **Eq. (9)** [47]:

$$\frac{1}{q}\left(\frac{dJp}{dx}\right) = G_{op}(x) - Rx \quad (9)$$

Where, $J_n$ and $J_p$ are denoted by the electron and hole current densities, respectively. The parameter *R(x)* is used for the net recombination rate that results from the band-to-band recombination and Shockley-Read-Hall (SRH) recombination traffic via the gap states. The total direct recombination rate is given in **Eq. (10)** [48]:

$$R_D(x) = \beta(np - ni^2) \quad (10)$$

Where, *n* and *p* are the existing band carrier concentrations when the devices are exposed to a voltage bias, light bias, or even both, and $\beta$ represents the proportionality constant based on the material's band structure under investigation. The term $G_{op}(x)$ is used for the optical generation rate as a function of *x* owing to the externally applied light, which is included in the continuity equation.

## 2.4 CsPbI$_3$-based PSC structure

The simulations were conducted for CsPbI$_3$-based PSC in this study that had an n-i-p planar heterojunction. The ETL, perovskite layer, and HTL were positioned in the n-region, i-region, and p-region, respectively. Exciton, a confined state consisting of an electron and a hole, is formed in the perovskite layer of the device when the cell is exposed to light. According to the length of their diffusion, they are capable to enter the n (p) area. The exciton splits apart at the interface between the n-layer and i-layer, sending the electrons and residual holes in the direction of the n-layer and p-layer, respectively. In a similar manner, the exciton splits apart at the interface between the i-layer and p-layer, causing the holes to go to the p-layer, while the residual electrons travel to the n-layer. The electrical field that exists between these layers helps to facilitate the dissociation of excitons, as well as the movement of electrons and holes.

Throughout the simulation, as the TCO, absorber, and back contact, the indium-doped tin oxide (ITO), CsPbI$_3$ perovskite (**Table 1**), and Au were employed, respectively. Furthermore, eight ETLs and twelve HTLs (**Table 1** and **Table 2**) were investigated for the PSC to explore the best combination. For each ETL in **Table 1**, we studied twelve different HTLs (**Table 2**) to get the best combination structures from 96 different combinations (**Figure 1(a)**). Additionally, the inclusion of interface defect layers is also supported by the SCAPS-1D software package. **Table 3** provides a summary of the defect density that was applied at each of the interfaces. Apart from evaluating the effect of working temperature on the device's efficiency, the simulations were conducted at a frequency and working temperature of 1 MHz and 300 K, respectively, under the AM1.5G 1 solar spectrum along with an incident power density of 1000 mW/cm$^2$.

**Table 1**. Input parameters of the ITO, ETLs, and absorber layer.

| Parameters | ITO | TiO$_2$ | PCBM | ZnO | C$_{60}$ | IGZO | SnO$_2$ | WS$_2$ | CeO$_2$ | CsPbI$_3$ |
|---|---|---|---|---|---|---|---|---|---|---|
| Thickness (nm) | 500 | 30 | 50 | 50 | 50 | 30 | 100 | 100 | 100 | 800* |
| Band gap, E$_g$ (eV) | 3.5 | 3.2 | 2 | 3.3 | 1.7 | 3.05 | 3.6 | 1.8 | 3.5 | 1.694 |
| Electron affinity, χ (eV) | 4 | 4 | 3.9 | 4 | 3.9 | 4.16 | 4 | 3.95 | 4.6 | 3.95 |
| Dielectric permittivity (relative), ε$_r$ | 9 | 9 | 3.9 | 9 | 4.2 | 10 | 9 | 13.6 | 9 | 6 |
| CB effective density of states, N$_C$ (1/cm$^3$) | 2.2 × 10$^{18}$ | 2 × 10$^{18}$ | 2.5 × 10$^{21}$ | 3.7×10$^{18}$ | 8.0 × 10$^{19}$ | 5 × 10$^{18}$ | 2.2 × 10$^{18}$ | 1 × 10$^{18}$ | 1 × 10$^{20}$ | 1.1 × 10$^{20}$ |
| VB effective density of states, N$_V$ (1/cm$^3$) | 1.8 × 10$^{19}$ | 1.8 × 10$^{19}$ | 2.5 × 10$^{21}$ | 1.8×10$^{19}$ | 8.0 × 10$^{19}$ | 5 × 10$^{18}$ | 1.8 × 10$^{19}$ | 2.4 × 10$^{19}$ | 2 × 10$^{21}$ | 8.2 × 10$^{20}$ |
| Electron mobility, μ$_n$ (cm$^2$/Vs) | 20 | 20 | 0.2 | 100 | 8.0 × 10$^{-2}$ | 15 | 100 | 100 | 100 | 25 |
| Hole mobility, μ$_h$ (cm$^2$/Vs) | 10 | 10 | 0.2 | 25 | 3.5 × 10$^{-3}$ | 0.1 | 25 | 100 | 25 | 25 |
| Shallow uniform acceptor density, N$_A$ (1/cm$^3$) | – | – | – | – | – | – | – | - | - | 1 × 10$^{15*}$ |
| Shallow uniform donor density, N$_D$ (1/cm$^3$) | 1×10$^{21}$ | 9 × 10$^{16}$ | 2.93 × 10$^{17}$ | 1 × 10$^{18}$ | 1 × 10$^{17}$ | 1 × 10$^{17}$ | 1 × 10$^{17}$ | 1×10$^{18}$ | 10$^{21}$ | 0* |
| Defect density, N$_t$ (1/cm$^3$) | 1 × 10$^{15*}$ | 1 × 10$^{15*}$ | 1 × 10$^{15*}$ | 1 × 10$^{15*}$ | 1 × 10$^{15*}$ | 1 × 10$^{15*}$ | 1 × 10$^{15*}$ | 1×10$^{15*}$ | 1 × 10$^{15*}$ | 1×10$^{15*}$ |
| References | 26,49,50 | 25,26 | 27–29,49,51 | 25,52 | 27 | 53,54 | 25,51 | 26,55 | 56 | 24 |

*In this study, these values remain constant during initial optimization to get the best combination of HTL, ETL, and back metal contact

**Table 2.** Input parameters of the HTLs.

| HTL | Cu$_2$O | CuSCN | CuSbS$_2$ | P3HT | PEDOT: PSS | Spiro-MeOTAD | NiO | CuI | CuO | V$_2$O$_5$ | CFTS | CBTS |
|---|---|---|---|---|---|---|---|---|---|---|---|---|
| Thickness (nm) | 50 | 50 | 50 | 50 | 50 | 200 | 100 | 100 | 50 | 100 | 100 | 100 |
| Band gap, E$_g$ (eV) | 2.2 | 3.6 | 1.58 | 1.7 | 1.6 | 3 | 3.8 | 3.1 | 1.51 | 2.20 | 1.3 | 1.9 |
| Electron affinity, $\chi$ (eV) | 3.4 | 1.7 | 4.2 | 3.5 | 3.4 | 2.2 | 1.46 | 2.1 | 4.07 | 4.00 | 3.3 | 3.6 |
| Dielectric permittivity (relative), $\varepsilon_r$ | 7.5 | 10 | 14.6 | 3 | 3 | 3 | 10.7 | 6.5 | 18.1 | 10.00 | 9 | 5.4 |
| CB effective density of states, N$_C$ (1/cm$^3$) | $2 \times 10^{19}$ | $2.2 \times 10^{19}$ | $2 \times 10^{18}$ | $2 \times 10^{21}$ | $2.2 \times 10^{18}$ | $2.2 \times 10^{18}$ | $2.8 \times 10^{19}$ | $2.8 \times 10^{19}$ | $2.2 \times 10^{19}$ | $9.2 \times 10^{17}$ | $2.2 \times 10^{18}$ | $2.2 \times 10^{18}$ |
| VB effective density of states, N$_V$ (1/cm$^3$) | $1 \times 10^{19}$ | $1.8 \times 10^{18}$ | $1 \times 10^{1}$ | $2 \times 10^{21}$ | $1.8 \times 10^{19}$ | $1.8 \times 10^{19}$ | $1 \times 10^{19}$ | $1 \times 10^{19}$ | $5.5 \times 10^{20}$ | $5.0 \times 10^{18}$ | $1.8 \times 10^{19}$ | $1.8 \times 10^{19}$ |
| Electron mobility, $\mu_n$ (cm$^2$/Vs) | 200 | 100 | 49 | $1.8 \times 10^{-3}$ | $4.5 \times 10^{-2}$ | $2.1 \times 10^{-3}$ | 12 | 100 | 100 | $3.2 \times 10^{2}$ | 21.98 | 30 |
| Hole mobility, $\mu_h$ (cm$^2$/Vs) | 8600 | 25 | 49 | $1.86 \times 10^{-2}$ | $4.5 \times 10^{-2}$ | $2.16 \times 10^{-3}$ | 2.8 | 43.9 | 0.1 | $4.0 \times 10^{1}$ | 21.98 | 10 |
| Shallow uniform acceptor density, N$_A$ (1/cm$^3$) | $1 \times 10^{18}$ | $1 \times 10^{18}$ | $1 \times 10^{18}$ | $1 \times 10^{18}$ | $1 \times 10^{18}$ | $1 \times 10^{18}$ | $1 \times 10^{18}$ | $1 \times 10^{18}$ | $1 \times 10^{18}$ | $1 \times 10^{18}$ | $1 \times 10^{18}$ | $1 \times 10^{18}$ |
| Shallow uniform donor density, N$_D$ (1/cm$^3$) | - | – | – | – | – | – | – | – | – | - | - | - |
| Defect density, N$_t$ (1/cm$^3$) | $1 \times 10^{15*}$ | $1 \times 10^{15*}$ | $1 \times 10^{15*}$ | $1 \times 10^{15*}$ | $1 \times 10^{15*}$ | $1 \times 10^{15*}$ | $1 \times 10^{15*}$ | $1 \times 10^{15*}$ | $1 \times 10^{15*}$ | $1 \times 10^{15*}$ | $1 \times 10^{15*}$ | $1 \times 10^{15*}$ |
| References | 25,57 | 25 | 25 | 25 | 58,59 | 25,26,60 | 61 | 62 | 63 | 64 | 35 | 35 |

*In this study

**Table 3.** Input parameters of the interface defect layers [25].

| Interface | Defect type | Capture cross-cection: Electrons/holes (cm$^2$) | Energetic distribution | Reference for defect energy level | Total density (cm$^{-2}$) (integrated over all energies) |
|---|---|---|---|---|---|
| ETL/CsPbI$_3$ | Neutral | $1.0 \times 10^{-17}$<br>$1.0 \times 10^{-18}$ | Single | Above the VB maximum | $1.0 \times 10^{10}$ |
| CsPbI$_3$/HTL | Neutral | $1.0 \times 10^{-18}$<br>$1.0 \times 10^{-19}$ | Single | Above the VB maximum | $1.0 \times 10^{10}$ |

## 3 Result and discussion

### 3.1 Analysis of DFT results

#### 3.1.1 Structural properties of CsPbI$_3$ compound

The crystal structure of the single-cubic-perovskite CsPbI$_3$ solar absorber with Pm3m space group (No. 221) is shown in **Figure 1(b)**. In the structure, there is a single formula unit with 5 atoms, where Cs is at the body center position, Pb occupies the corner position, and I is at the face center of the cube. The calculated lattice parameter of the best-optimized structure of CsPbI$_3$ was 6.407 Å, which excellently matches the experimental data (6.414 Å). Very consistent lattice parameters and negative formation energy (-6.905 eV/atom) indicate the cell stability of the solar cell absorber material as well.

#### 3.1.2 Band structure and DOS of CsPbI$_3$ solar compound

The electronic band diagram and corresponding density of states (DOS) of CsPbI$_3$ cubic perovskite towards the better symmetry directions (X-R-M-Γ-R) of the Brillouin zone are presented in **Figures 1(c)** and **1(d)**, respectively. The Fermi level ($E_F$) in these figures is shown by a horizontal dotted line. It is well recognized that photon-absorbing materials (solar cells) always exhibit band gap ($E_g$) energy that belongs to the class of semiconducting substances. From **Figure 1(c)**, it is obvious that there is a direct band gap, $E_g$ ~1.48 eV for the CsPbI$_3$ while applying the GGA-PBE method, which is somewhat lower than the previous report (1.694 eV)[24]. This discrepancy in band gap energy might be originated because of the utilization of GGA-PBE potential (average potential) that generally debates the band gap of semiconductors[65,66]. **Figure 1(d)** depicts the total density of states (TDOS) and partial density of states (PDOS) that show the contributions of different orbital electrons, as well as the chemical bonding nature of CsPbI$_3$. The shown TDOS curve indicates that the compound has n-type carriers with a small value of TDOS (1.98 states/eV) at $E_F$ owing to its semiconducting nature. The occurred individual band PDOS from Pb-5$d$ orbital electrons contributes mostly in contrast to any other orbital electrons at $E_F$, whereas the Cs-6$s$ orbital has a moderate contribution, and I-5$p$ has a trivial effect in the vicinity of $E_F$ for the generation of carriers. Therefore, amid the PDOS of CsPbI$_3$, the Pb atom orbital electrons play key roles in the absorption of photon energy and hence the generation of photocurrent.

#### 3.1.3 Electron charge density of CsPbI$_3$ compound

**Figure 1(e)** displays the charge density map toward the (100) crystallographic plan, where the scale bar on the right side implies the intensity of electron density according to colors. It is observed that the charges gather dominantly around the Pb atom, and the charge ordering around Pb is identic also. The Pb-I bond is punchy owing to the hybridization between Pb-5$d$ and I-5$p$ orbitals, which is also seen from the DOS curve. Nonetheless, the Cs atom in the charge density figure is not visualized separately because of the charge overlapping of Cs with the I-atom. Moreover, the charge distribution map revealed that the Cs and I ions create an ionic bond, while the covalent bond was found for the Pb and I ions. This covalent nature of Pb and I ion is used to create a weaker bond, whereas the ionic bond increases the bonding nature in the structural unit. In this structure, the covalent bond occurred due to the hybridization of Pb-5$d$ and Cs-6$s$ states[67,68]. It is found that the charge distribution is almost spheric around the entire atoms, which is a signature of ionic bonding that can be compared with reported perovskites[69–71] as well. Furthermore, the electronic charge density maps along different crystal planes originated identical results, which is an indication of its isotropic behavior.

#### 3.1.4 Fermi surface of CsPbI$_3$ compound

The computed Fermi surface topology at different paths of the Brillouin zone of the CsPbI$_3$ solar absorber is depicted in **Figure 1(f)**. It is visualized that at the R(0.5,0.5,0.5) point of each corner, a separated small hole-like rectangular curve shape sheet is connected with a circular disc/plate-like electron sheet at X(0.5,0.0,0.0) point of body center. There are six electron-like sheets that are found on six faces of the cube. Moreover, six hole-like spheroid giant hollow sheets are also present at the body centers by covering the path R-M-Γ-R, where the R-point is directly linked with the center Γ(0.0,0.0,0.0) point in the topology of the Fermi surface. Therefore, both electron-like and hole-like Fermi surfaces exist, which indicates the multi-band feature of the titled material.

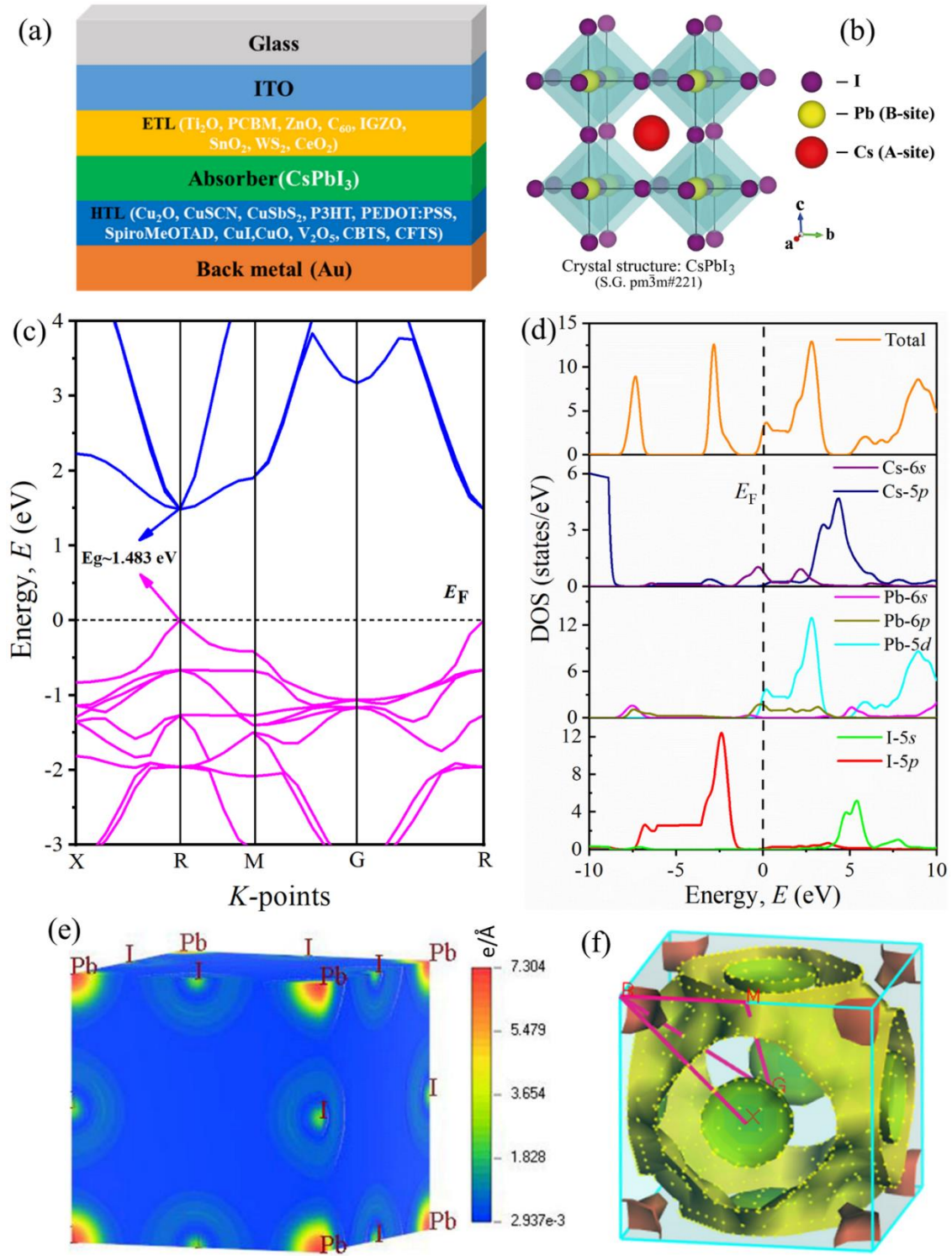

**Figure 1**. (a) Design configuration of the CsPbI$_3$-based PSC, (b) the crystal structure of CsPbI$_3$ cubic-single-perovskite semiconductor, (c) computed band structure along noted symmetry direction of Brillouin zone using GGA-PBE approach, (d) total and partial density of states of CsPbI$_3$ semiconductor, (e) electronic charge density map along (100) plane, and (f) Fermi surface topology of CsPbI$_3$ semiconductor perovskite.

### 3.1.5 *Optical properties of CsPbI$_3$ compound*

The optical functions are very important to study the electronic structure and potential optical applications in the arena of optoelectronics as well as nano-electronic devices. Therefore, in this study, we calculated frequency-dependent six optical parameters such as dielectric function ($\varepsilon(\omega)$), electron energy loss function (L($\omega$)), refractive index (n($\omega$)), optical absorption ($\alpha(\omega)$), reflectivity (R($\omega$)), and conductivity ($\delta(\omega)$). The calculated optical characteristics of CsPbI$_3$ solar absorber material and its architectures were investigated from the energy range of 0 to 40 eV, where the real ($\varepsilon_1$) and imaginary ($\varepsilon_2$) parts are shown by solid and broken/dotted lines, respectively, in **Figures 2(a)-(c)**.

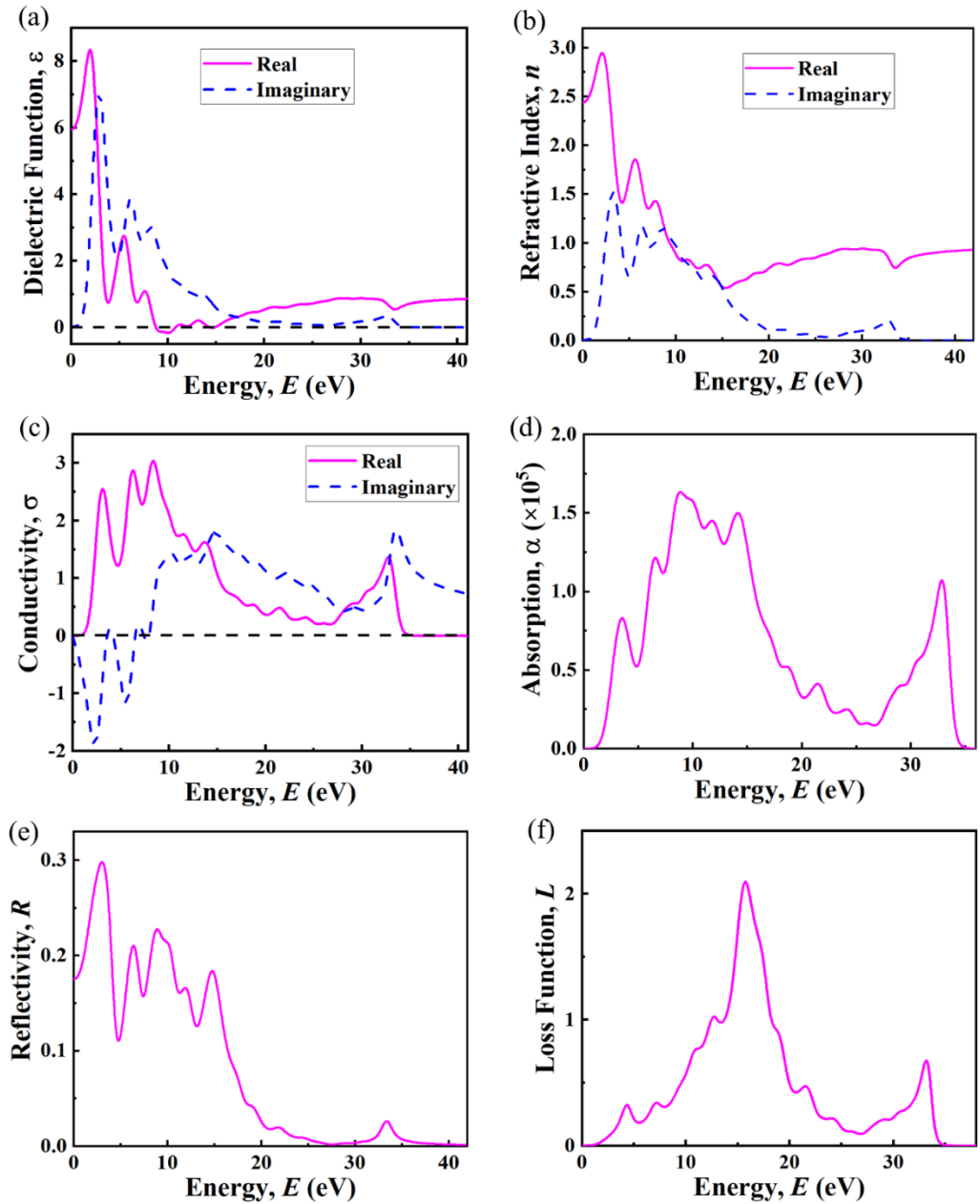

**Figure 2**. Frequency-dependent optical properties of CsPbI$_3$: (a) dielectric function ($\varepsilon$), (b) refractive index ($n$). Frequency-dependent optical properties of CsPbI$_3$: (c) photoconductivity ($\sigma$), (d) absorption coefficient ($\alpha$), (e) reflectivity ($R$), and (f) loss function ($L$).

**Figure 2(a)** represents the real ($\varepsilon_1$) and imaginary ($\varepsilon_2$) spectra in terms of the photon energy of CsPbI$_3$, where $\varepsilon_1$ of real dielectric function at zero energy is regarded as the electronic portion of static dielectric function that has the exactly 6.0 value at zero frequency. The $\varepsilon_1$ gradually increases with the increment of photon energy and reaches the apex value of 8.26 at around 2.64 eV. After that, it sharply drops to a lower dielectric function and again rises at 4.96 eV. Next, another peak is seen at 7.54 eV. However, $\varepsilon_1$ reaches downs to zero at 8.86 eV and returns to zero again at around 10.56 eV. The observed negative dielectric function in this frequency range is an indication of the Drude-like behavior of the material. In addition, the dielectric function is related to intra and inter band transitions, where intra band transition generally occurs due to the absorption of free or conduction electrons for metal systems. There are three main observed peaks at 2.64 eV, 4.96 eV, and 7.54 eV in the investigated frequency region. Among these, the highest peak at 2.64 eV in the visible region originated owing to the inter band transitions of electrons, whereas the other two low-intensity peaks are found for the lower intensities of semi-core states of the conduction bands. From 20.12 eV to 32.68 eV, the dielectric function exhibits an almost linear response with an increase in frequency. There was a certain amount of energy loss due to the electrons. In this range of frequency, the energy loss originates due to the electron transitions, and the dielectric function becomes almost zero or negative. Thus, no wave propagated from this region [72]. Notably, the imaginary dielectric function, $\varepsilon_2(\omega)$ attains zero in the ultraviolet spectrum at about 25 eV, which proves the transparent character and optically anisotropic nature of $CsPbI_3$ as well.

The parameters refractive index (*n*) and extinction coefficient (*k*) deal with the absorption loss, as well as the changes in the phase velocity during the propagation of the electromagnetic signal (light) throughout the compounds. The computed frequency-dependent *n* and *k* spectra are displayed in **Figure 2(b)**, where *n*(0) = 2.45 is seen for the $CsPbI_3$ solar absorber [72]. We notice that the refractive index increases with the increment of radiative light (2.08 eV), and the highest value of the refractive index was 2.98. Moreover, at 5.63 eV and 7.87 eV energies, the refractive indexes were 0.87 and 1.43, respectively, meaning that the refractive index varies with the external frequency, which is an indication of the photorefractive properties of $CsPbI_3$. However, the found value of *n*(0) of the material is comparable with the $K_2Cu_2GeS_4$ and GaAs semiconductors [73].

**Figure 2(c)** shows the spectra of photoconductivity, where the changes of electrical conductivity ($\sigma$) are very analogous to that of absorption coefficient ($\alpha$). It is seen from **Figure 2(c)** that the real portion of the $\sigma$ commences from zero frequency, though $CsPbI_3$ is a moderate band gap (1.489 eV) semiconductor as seen from the band diagram of **Figure 1(d)**. The aforementioned phenomenon might be originated due to a lack of more accurate potentials and the degenerate semiconducting nature or high electrical conducting behaviour of Pb-6*p*/5*d* and Cs-6*s* orbital electrons at $E_F$ (**Figure 1(d)**). Thus, it is uttered that the studied material behaves like a degenerate semiconductor, which is a convenient feature for high-efficiency solar cell applications. However, in the case of photoconductivity, three prominent peaks appeared in the range of 4.5 eV to 9.8 eV, where the highest magnitude (3.03) peak is observed at 9.79 eV. After reaching the maximal value, the photoconductivity tends to reduce with photon energy.

The optical function, α shows the number of photons absorbed by the substances, as well as provides important data about the conversion efficiency of solar energy [74]. **Figure 2(d)** reveals the *α* curve of $CsPbI_3$ that is involved with the optimal solar energy altering efficiency and initiates with approximately 1.48 eV because of the semiconducting character along with the predicted band gap, as well as visualized from the band diagram. Commonly, it is observed that the peaks in the low energy infrared range originate from the intra band transition [72], whereas the peaks in the exalted-frequency level of absorption and conductivity spectra are seen forasmuch as the inter band transitional nature. It is seen that three main absorption peaks arise in the extent of 7 eV to 14 eV, which originate due to the absorption of photons from the ultraviolet energy region. Among these, the utmost absorption is discovered at 8.64 eV owing to the intra band transition. In contrast, other high-level intensity peaks from 18 eV to 30 eV arise for the inter band electronic transition rates. Importantly, the *σ*, and hence, the photoconductivity of a semiconductor material raises with absorbed lights [75] and exhibits a similar trend like the α function.

**Figure 2(e)** represents the frequency-dependent reflectivity spectrum (*R*) for $CsPbI_3$, which is an important optical function for the determination of all optical constants by employing the Kramers-Kroning relations. From **Figure 2(e)**, we see that the *R*-value started at zero energy is considered as the static part of reflectivity, where the highest reflectivity (0.3) is visualized in the infrared region (3.85 eV) for the intra band transition in the compound. Moreover, the *R*(0) = 1.75 is observed for this solar cell absorber. Furthermore, few peaks with low intensity are found for it in the ultraviolet range owing to the inter band transitions. We also noticed that the reflectivity declines with the increment of the energy band gap [72,76] in the material.

**Figure 2(f)** shows the energy loss spectrum (*L*) of $CsPbI_3$, which indicates how much energy a fast electron loses during propagating through a compound. The maximal loss function (*L*) is associated with the plasma resonance, and its consorted frequency is called the plasma frequency, $\omega_p$ [77]. The highest peak value (2.1) for this compound is observed at 15.91 eV, which indicates the plasma frequency of the perovskite. It is seen that the main energy loss occurs in the ultraviolet region due to the higher photon energy than the band gap energy [72]. However, the relatively low value of $\omega_p$ and $L$ might be originated because of the semiconducting behavior and/or the bigger effective mass of the conduction electrons.

## 3.2 Analysis of SCAPS-1D results

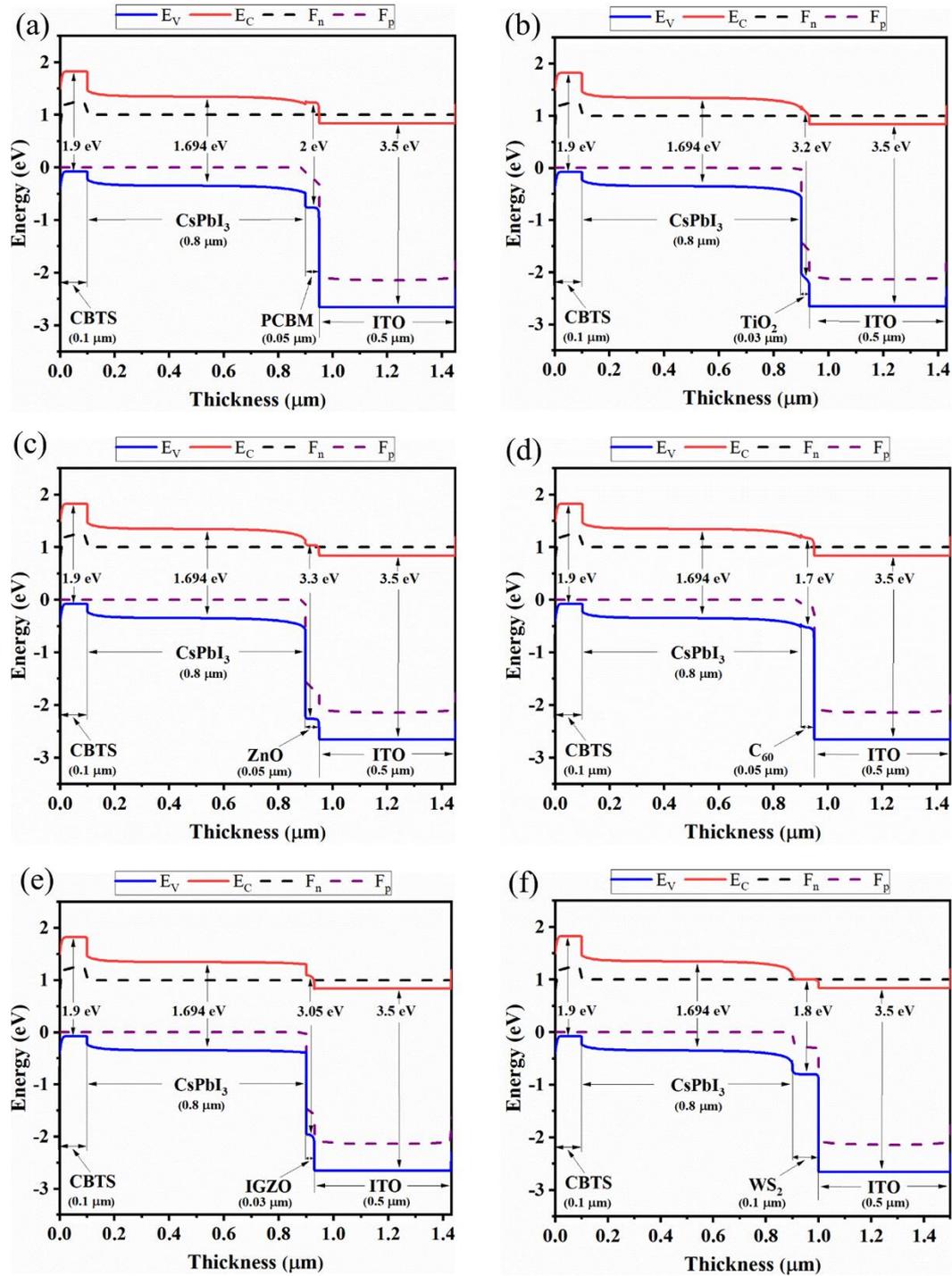

**Figure 3.** Energy diagram of CsPbI$_3$-based PSCs with CBTS as the HTL and (a) PCBM, (b) TiO$_2$, (c) ZnO, (d) C$_{60}$, (e) IGZO, and (f) WS$_2$ as the ETL.

### 3.2.1 Band diagram and energy level alignment

**Figure S1** shows the energy level alignment of the studied CsPbI$_3$ absorber, ITO, ETLs, and HTLs, and **Figure 3** shows the energy diagram of CsPbI$_3$-based best optimized six (06) PSCs. To extract the electron at the ETL/CsPbI$_3$ interface, the electron affinity of ETL must be greater than that of CsPbI$_3$, whereas the ionization energy of CBTS HTL must be less than that of CsPbI$_3$ to extract the holes at the CsPbI$_3$/HTL interface. Additionally, the energy band imbalance at both interfaces has a considerable impact on the performance parameters of the device (**Figure 3**). In **Figure 3**, each device's quasi-Fermi levels, $F_n$ and $F_p$, cohabited with conduction band energy ($E_C$) and valence band energy ($E_V$), respectively. In each ETL, the $F_p$ was superimposed over the $E_V$, while the $F_n$ and $E_C$ carried on in a manner that was harmonically comparable. The bandgap of $CsPbI_3$ was 1.694 eV, whereas the bandgaps of PCBM, $TiO_2$, ZnO, $C_{60}$, IGZO, and $WS_2$ ETLs were 2 eV, 3.2 eV, 3.3 eV, 1.7 eV, 3.05 eV, and 1.8 eV, respectively (**Figure S1**). Due to the similar nature of band alignment, the performance of the $TiO_2$ and ZnO ETLs was relatively similar. On the other hand, it was discovered that the $F_n$ was able to pass through the $E_C$, but the $F_p$ and $E_V$ continued to exist at the same level in the CBTS HTL compared to other HTLs (**Figure 3**). This kind of behavior prevents the holes from moving into the ETL and electrons from moving toward the HTL. As a direct consequence of this, the rear contact is able to readily collect holes from the CBTS HTL, while the front contact is able to simply collect electrons from the $CsPbI_3$ absorber layer. As a result of this idea, the rear contact was made of gold, which had a work function of 5.1 eV, while the TCO was made of indium dioxide, which held a work function of 4.0 eV. On the other hand, PCBM, $C_{60}$, and IGZO ETLs showed relatively lower performance because of their corresponding bandgaps (**Figure S1**).

### 3.2.2 Effect of ETL variation

The performance of $CsPbI_3$-based PSC was optimized by employing different configurations of 12 HTLs and 8 ETLs while keeping Au as the back contact. However, as $SnO_2$ and $CeO_2$ ETLs showed very poor performance, they were disregarded while plotting the performance graphs for different combinations as shown in **Figure 4**.

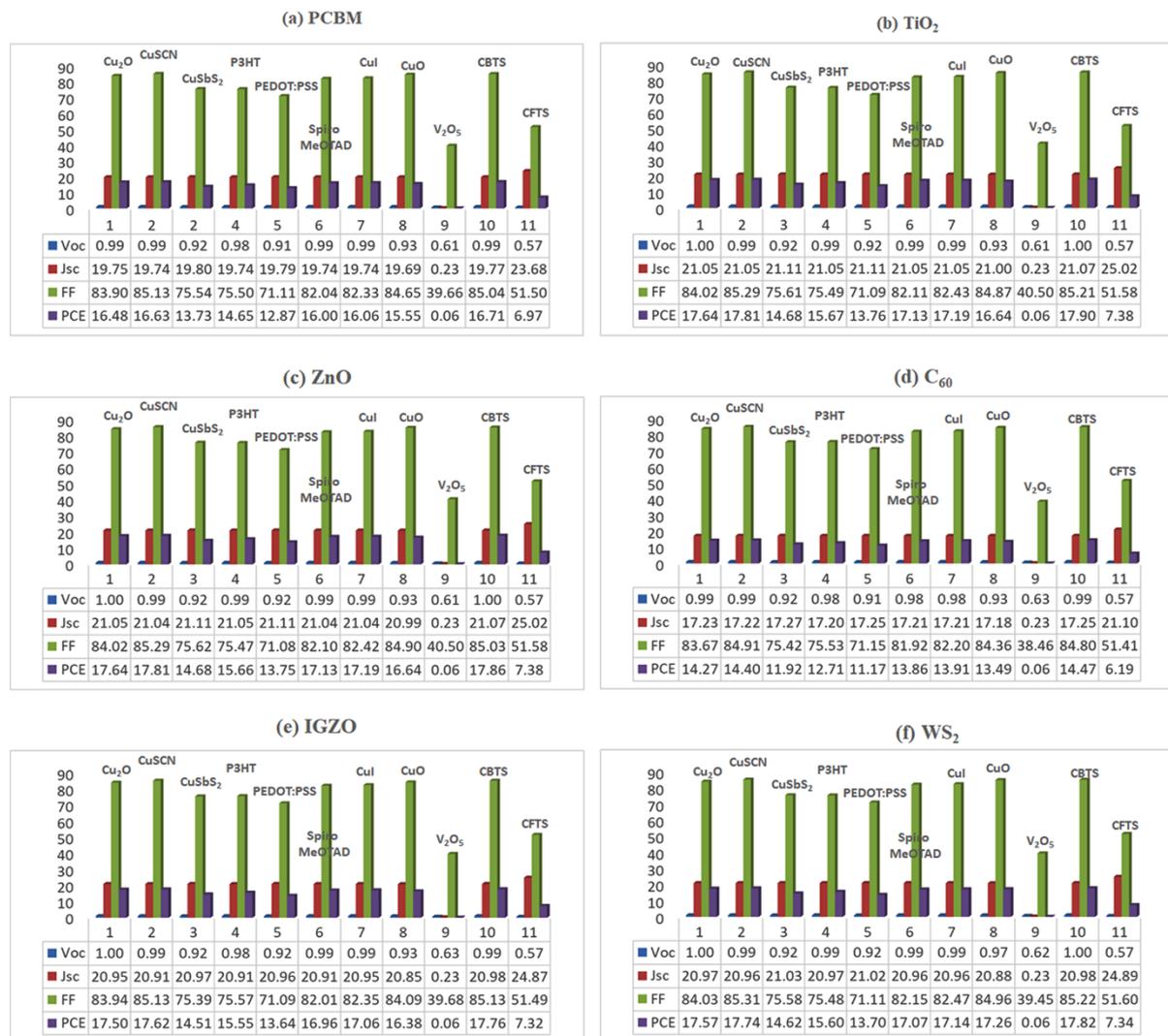

**Figure 4.** Performance of $CsPbI_3$-based PSC for various HTLs with Au as the back contact and (a) PCBM, (b) $TiO_2$, (c) ZnO, (d) $C_{60}$, (e) IGZO, and (f) $WS_2$ as the ETL.

During the optimization, the absorber thickness, acceptor density, donor density, and defect density were kept constant at 800 nm, $10^{15}$ cm$^{-3}$, 0 cm$^{-3}$, and $10^{15}$ cm$^{-3}$, respectively, for finding out the best configuration of ETL and HTL. In addition, for each HTL, ETL, and ITO, the defect density was $10^{15}$ cm$^{-3}$. The best 6 configurations of PSC to achieve the highest PCE by using a variety of ETLs with the best HTL are mentioned in **Table 4**. **Figure 4** and **Table 4** illustrate that the ITO/TiO$_2$/CsPbI$_3$/CBTS/Au configuration achieved the highest PCE of 17.90%, with a short circuit current density ($J_{SC}$) of 21.075 mA/cm$^2$, field factor (*FF*) of 85.21%, and open circuit voltage ($V_{OC}$) of 0.997 V. From **Figure 4**, we found that ETLs like TiO$_2$, ZnO, WS$_2$, and IGZO achieved nearly 18% *PCE* due to their band alignment in comparison to other ETLs such as C$_{60}$, PCBM, CeO$_2$, and SnO$_2$ (**Figure S1**). When being employed with TiO$_2$ ETL, the HTLs like Cu$_2$O, CuSCN, and CBTS achieved significant *PCEs* of 17.64%, 17.81%, and 17.90%, respectively, whereas ZnO and WS$_2$ ETLs achieved similar *PCEs* (>17.5%) while being used with those HTLs. From the energy diagrams (**Figure 3**), it was observed that TiO$_2$ and ZnO ETLs were positioned closer, and the bandgap of TiO$_2$ and ZnO was 3.2 eV and 3.3 eV, respectively. Because of that, the configurations with TiO$_2$ and ZnO as the ETL showed very close PCE of 17.9% and 17.86%, respectively. In a few studies, both of them showed an efficiency of around 25% when used as the ETL [78]. To find better-optimized results, the thickness, donor density, and defect density of these ETLs were optimized in the previous findings enlisted in **Table 1** [25,26,52]. After considering all the aspects, 6 best-performed ETLs were considered for the onward study (**Table 4**).

**Table 4:** Performance parameters for the combinations of different ETLs with the best HTL.

| Optimized Device | *Voc* (V) | *Jsc* (mA/cm$^2$) | FF (%) | PCE (%) |
| --- | --- | --- | --- | --- |
| ITO/PCBM/CsPbI$_3$/CBTS/Au | 0.994 | 19.77 | 85.04 | 16.71 |
| ITO/TiO$_2$/CsPbI$_3$/CBTS/Au | 0.997 | 21.07 | 85.21 | 17.90 |
| ITO/ZnO/CsPbI$_3$/CBTS/Au | 0.997 | 21.07 | 85.03 | 17.86 |
| ITO/C$_{60}$/CsPbI$_3$/CBTS/Au | 0.990 | 17.25 | 84.80 | 14.47 |
| ITO/IGZO/CsPbI$_3$/CBTS/Au | 0.995 | 20.98 | 85.13 | 17.76 |
| ITO/WS$_2$/CsPbI$_3$/CBTS/Au | 0.997 | 20.98 | 85.22 | 17.82 |

### 3.2.3 *Effect of HTL variation*

When eight sets of ETLs were optimized with twelve separate HTLs, compared to the other HTLs, the CBTS, CuSCN, CuI, and Cu$_2$O HTLs showed superior *PCEs* of around 17% to 18% for all the ETLs except C$_{60}$, CeO$_2$, and SnO$_2$ ETLs. When viewed from the perspective of the band diagram, these HTLs seemed to be most suited for the CsPbI$_3$ absorber. On the other hand, in contrast to the other HTLs, the V$_2$O$_5$ and CFTS HTLs showed lower *PCEs* of 0.6% and <8%, respectively, when paired with each ETL (**Figures S1 and 4**). As HTL is a p-type layer, it has to be thicker in contrast to the n-type ETL to lower the possibility of recombination. Because this permits the quick transfer of an equal number of charge carriers to the device [79]. As a consequence, the thickness of each ETL was smaller in comparison with the corresponding HTL. When we tested the inorganic HTLs, we found that they performed better than the organic HTLs in most cases with some exceptions. This is because the inorganic HTLs have features such as increased stability, high transparency, and good band alignment. For each case of **Figure 4**, we found that CBTS HTL, an earth-abundant material, showed excellent results for each ETL due to its excellent crystalline structure, light-absorbing ability, and atomic size [35,37,38]. **Figure S2** illustrates the best-optimized structures of CsPbI$_3$-based PSC with CBTS as the HTL and PCBM, TiO$_2$, ZnO, C$_{60}$, IGZO, and WS$_2$ as the ETLs.

### *3.2.4 Effect of absorber and ETL thickness on the device performance*

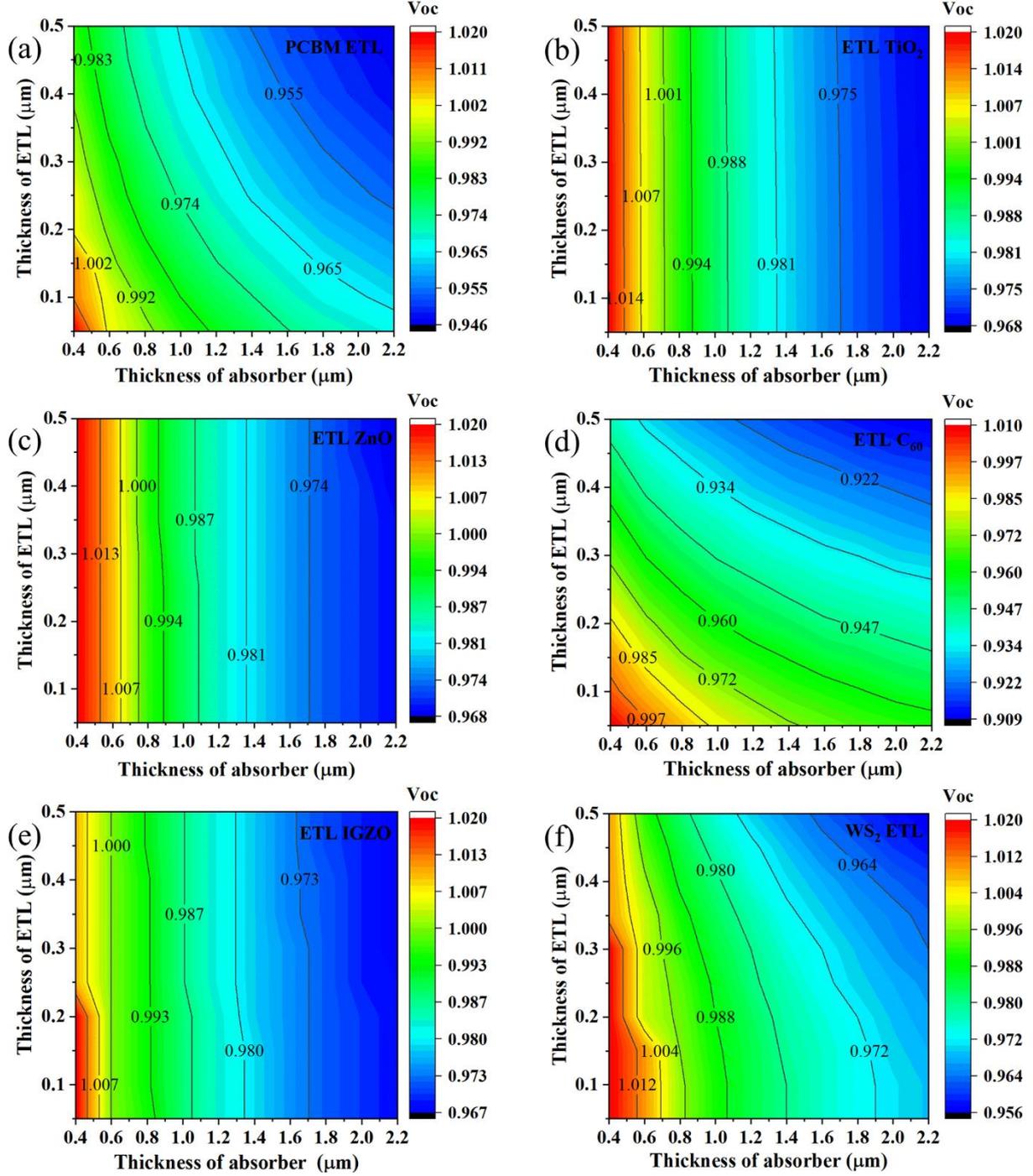

**Figure 5**: Contour mapping of $V_{OC}$ with respect to the thickness of CsPbI$_3$ absorber and (a) C$_{60}$, (b) IGZO, (c) PCBM, (d) TiO$_2$, (e) WS$_2$, (f) ZnO ETLs, respectively.

The contour maps of the projected $J_{SC}$, $V_{OC}$, $FF$, and efficiency with respect to the varying ETL thickness (50 nm to 500 nm) and absorber thickness (400 nm to 2200 nm) have been shown in **Figures** (5-8) for the CsPbI$_3$-based PSCs. In **Figure 5**, when the absorber and ETL thickness was around 400 nm and 50 nm, respectively, the PSCs with PCBM, TiO$_2$, ZnO, IGZO, and WS$_2$ ETLs showed the maximum $V_{OC}$ of 1.02 V, while the PSC with C$_{60}$ ETL showed the maximum $V_{OC}$ of 1.01 V. With the increase of absorber and ETL thickness, the $V_{OC}$ of each device declined because of the rise of the reverse saturation current due to the presence of a thicker absorber and the partial absorption of light by a thicker ETL. Among all the devices, the lowest $V_{OC}$ of 0.909 V was found for the PSC with C$_{60}$ ETL when the absorber and ETL thickness was 2200 nm and 500 nm, respectively.

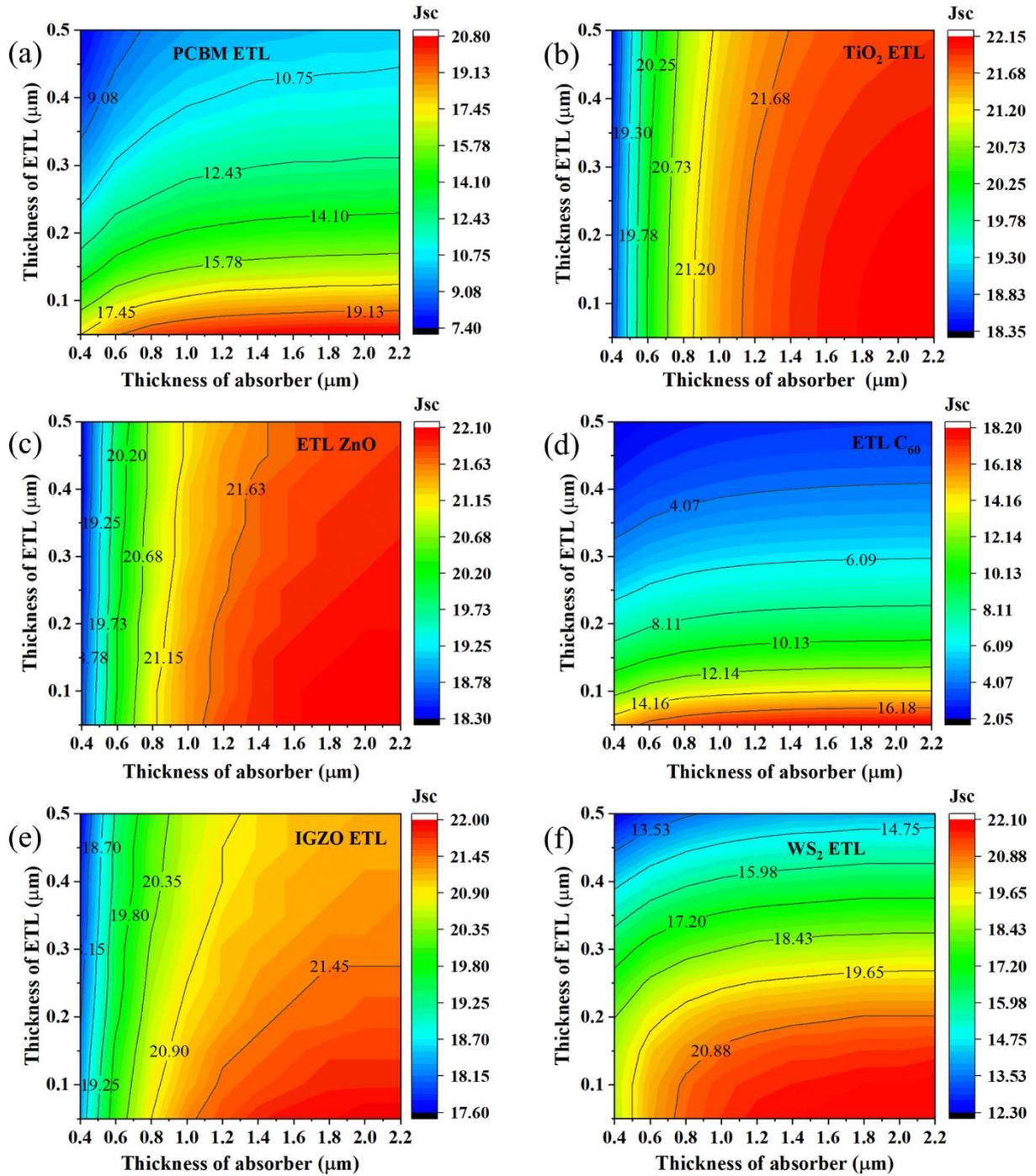

**Figure 6.** Contour mapping of $J_{SC}$ with respect to the thickness of CsPbI$_3$ absorber and (a) C$_{60}$, (b) IGZO, (c) PCBM, (d) TiO$_2$, (e) WS$_2$, (f) ZnO ETLs, respectively.

In the case of the $J_{SC}$ (**Figure 6**), the PSC with TiO$_2$ ETL showed a maximum value of 22.15 mA/cm$^2$ when absorber and ETL thickness was ≥ 1800 nm and ≤ 250 nm, respectively. The PSCs with ZnO and WS$_2$ ETLs showed the maximum $J_{SC}$ values of 22.10 mA/cm$^2$ when the absorber and ETL thickness was ≥ 1400 nm and ≤ 150 nm, respectively. Besides, the PSCs with IGZO, PCBM, and C$_{60}$ ETLs exhibited the maximum $J_{SC}$ values of 22.00, 20.80, and 18.20 mA/cm$^2$, respectively, when the absorber and ETL thickness was between 1600-2200 nm and 50-100 nm, respectively. Generally, for each device, the $J_{SC}$ increased with the increment of absorber thickness due to the rise of spectral response at longer wavelengths, while the enhancement of ETL thickness lowered the $J_{SC}$ due to the partial absorption of light.

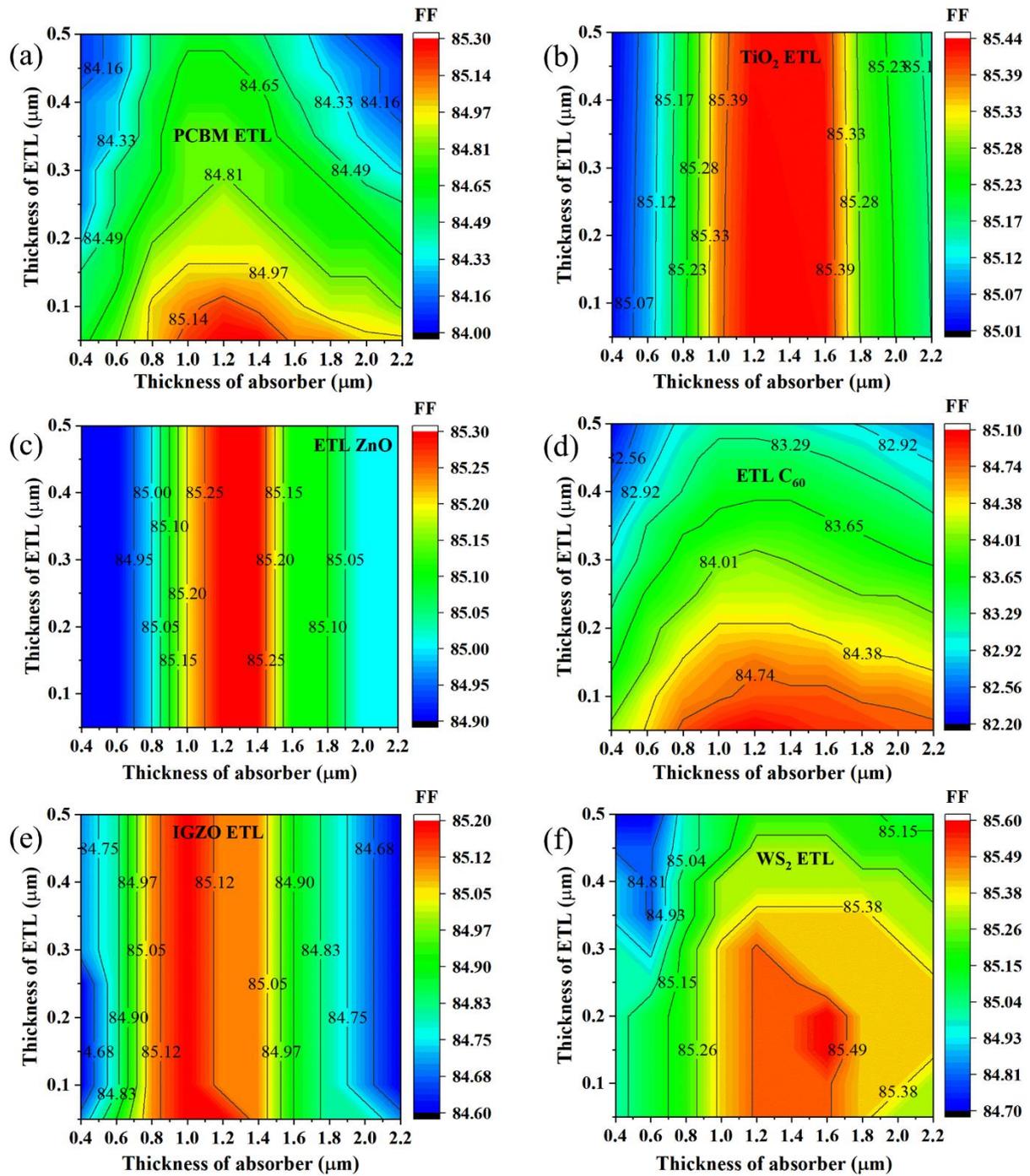

**Figure 7**: Contour mapping of *FF* with respect to the thickness of CsPbI$_3$ absorber and (a) C$_{60}$, (b) IGZO, (c) PCBM, (d) TiO$_2$, (e) WS$_2$, (f) ZnO ETLs, respectively.

In **Figure 7**, among all the CsPbI$_3$-based PSCs, the WS$_2$ ETL showed the maximum *FF* of 85.65% between 1400-1600 nm and 100-200 nm of absorber and ETL thickness, respectively. For the TiO$_2$ ETL, a more stable *FF* of 85.44% was found when the absorber thickness was between 1000-1600 nm, and the ETL thickness was between 100-400 nm. For the ETLs like PCBM and ZnO, a similar *FF* of 85.30% was obtained while the absorber thickness ranged from 1000 nm to 1400 nm, and the ETL thickness was ≤ 200 nm. The PSC with C$_{60}$ ETL exhibited the minimum FF of about 85.10% between 1000-1400 nm and 50-100 nm of absorber and ETL thickness, respectively.

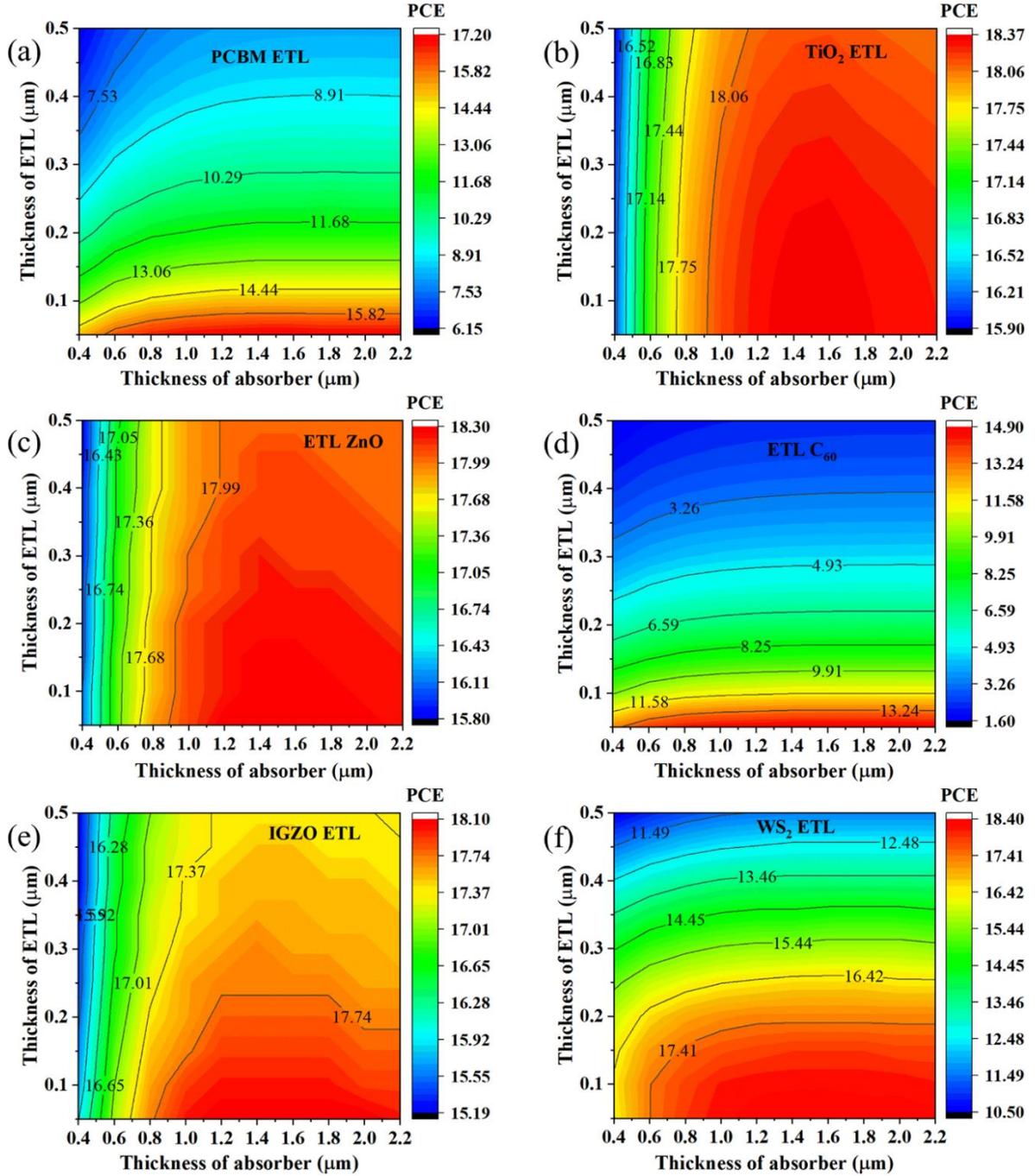

**Figure 8**: Contour mapping of PCE with respect to the thickness of CsPbI$_3$ absorber and (a) C$_{60}$, (b) IGZO, (c) PCBM, (d) TiO$_2$, (e) WS$_2$, (f) ZnO ETLs, respectively.

In **Figure 8**, among all the devices, the PSC with WS$_2$ ETL showed the highest efficiency of 18.40% between 1000-2000 nm absorber thickness and 50-100 nm ETL thickness. However, the devices with TiO$_2$ and ZnO ETLs also showed a good and consistent PCE of 18.37% and 18.30%, respectively, between 1200-2000 nm absorber thickness and 100-250 nm ETL thickness. In comparison with other devices, the PSC with C$_{60}$ ETL showed the lowest PCE of 14.90% with an absorber thickness between 1400 nm to 1600 nm and an ETL thickness of around 50 nm. As the PCE is a function of $J_{SC}$, $V_{OC}$, and $FF$ parameters, with the increment of absorber and ETL thickness, the PV performance increased up to a certain range depending on the absorber and ETL material of the PSC parallel to the previous study [80].

### *3.2.5    Effect of series resistance*

**Figure 9** represents the effect of series resistance ($R_S$) in the ranges of 0-6 $\Omega\text{-cm}^2$ on the performance of ITO/ETL/CsPbI$_3$/CBTS/Au structure, whereas the shunt resistance ($R_{SH}$) was kept constant at $10^5$ $\Omega\text{-cm}^2$. From the figures, it could be identified how the variation of $R_S$ affected the $V_{OC}$, $J_{SC}$, $FF$, and $PCE$. With the increasing $R_S$ from 0-6 $\Omega\text{-cm}^2$, the $V_{OC}$ increased marginally for all the studied devices, while the $J_{SC}$ of every device was almost unchanged. On the other hand, both the $FF$ and PCE decreased significantly by 11.02% and 2.24% on average (in magnitude), respectively, with the increment of $R_S$ for each device. As a non-ohmic contact in a PSC shows a high $R_S$, the highly conductive ITO layer rarely induces any $R_S$ [81]. The $R_S$ in a solar cell is made up of multiple resistances, including the resistance at the interface between semiconductors and metal contacts. Even though there is no net flow of current through the $R_S$, it might sometimes affect the $V_{OC}$ [50]. If the $R_S$ is raised, there is a significant reduction in the $FF$, which also decreases the PCE because of the higher power loss [24]. For this reason, with the increment of $R_S$, the $FF$ and PCE decreased in this work.

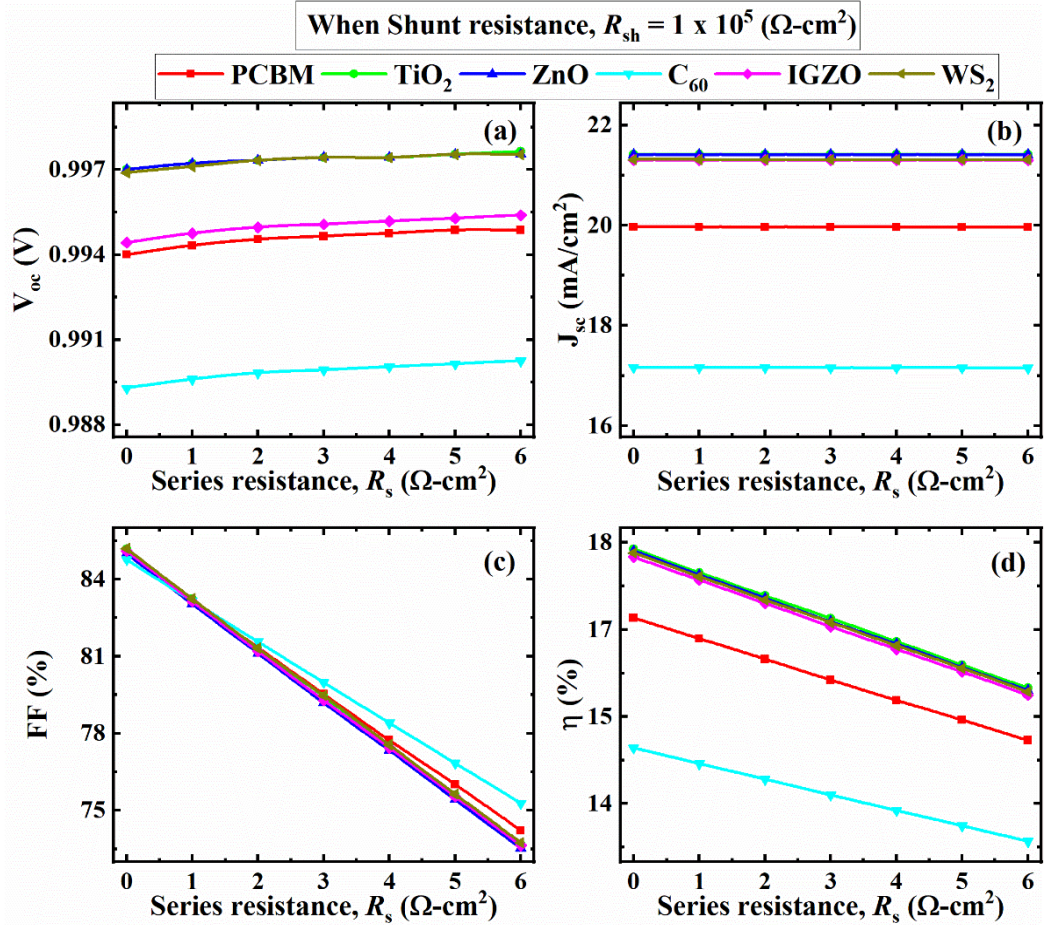

**Figure 9.** Effect of series resistance ($R_S$) on (a) $V_{OC}$, (b) $J_{SC}$, (c) $FF$, and (d) PCE when $R_{SH} = 10^5$ $\Omega\text{-cm}^2$.

### *3.2.6    Effect of shunt resistance*

**Figure 10** illustrates the effect of the alteration in shunt resistance ($R_{SH}$) in the ranges of 10 $\Omega\text{-cm}^2$ to $10^7$ $\Omega\text{-cm}^2$ on the $V_{OC}$, $J_{SC}$, $FF$, and PCE of each device, while the $R_S$ remained constant at 0.5 $\Omega\text{-cm}^2$. The $V_{OC}$ and $J_{SC}$ increased up to $10^2$ $\Omega\text{-cm}^2$ $R_{SH}$ for each of the six studied devices. After that, they were unchanged from $10^3$ $\Omega\text{-cm}^2$ to $10^7$ $\Omega\text{-cm}^2$. On the other hand, the $FF$ and PCE increased by 55.51% and 15.16% on average (in magnitude), respectively, from $10\text{-}10^3$ $\Omega\text{-cm}^2$ $R_{SH}$ values for each of the studied devices, and then, both became stable. The defects that occur during the manufacturing of the device give rise to $R_{SH}$ [56]. The results of our research show that the $R_{SH}$ can significantly improve the $FF$ and PCE. A greater $R_{SH}$ transforms the p-n junction into a route with low resistance so that the junction current may flow more easily, which can improve the device's performance [50,56].

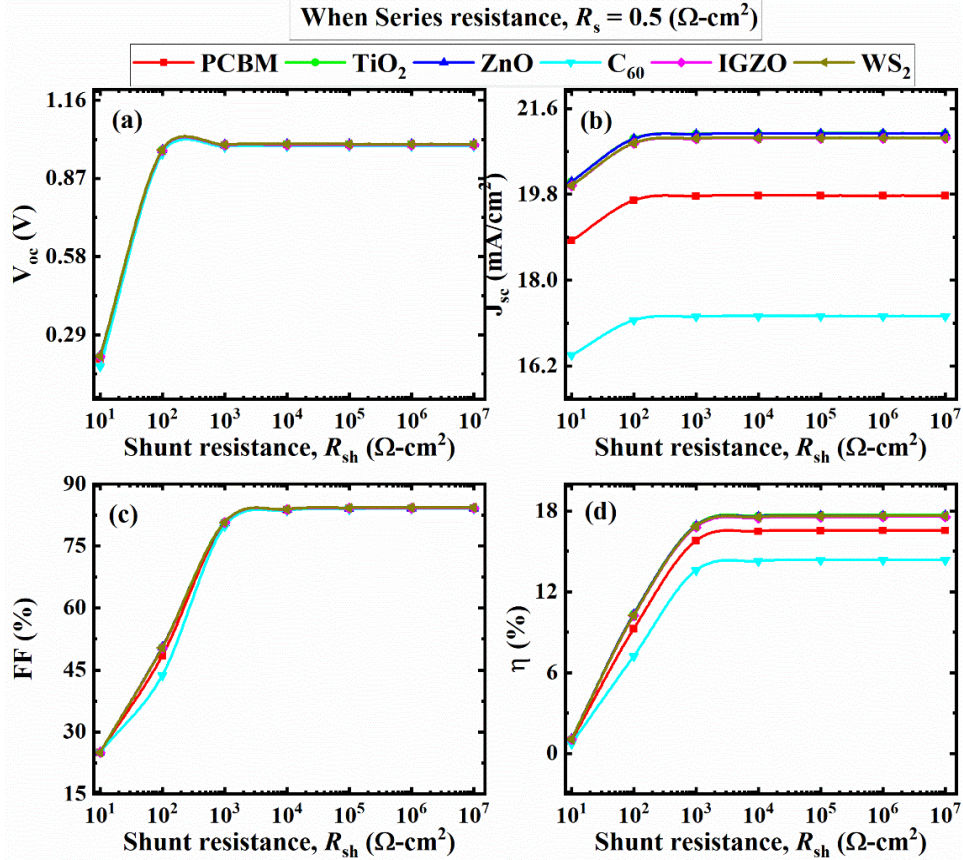

**Figure 10.** Effect of shunt resistance ($R_{SH}$) on (a) $V_{OC}$, (b) $J_{SC}$, (c) $FF$, and (d) PCE when $R_S$ = 0.5 Ω-cm$^2$.

### 3.2.7 Effect of temperature

**Figure S3** depicts the influence of the change of temperatures from 275 K to 450 K on the electrical outputs like $V_{OC}$, $J_{SC}$, $FF$, and PCE of the six devices. During the study, we noticed that the $V_{OC}$, $FF$, and PCE decreased with the increase of the temperature for every structure, whereas the $J_{SC}$ changed slightly for every device. The steady drop in the device performance with a temperature rise is in line with the findings that were obtained before [82]. This could be a result of the defect density inside the layers increasing with temperature, causing an efficiency loss. In addition, when the temperature rises, the deformation stress also moves onward, which in turn, decreases the efficiency by 10.33% on average (in magnitude). The rising temperature influences the diffusion length, and as a consequence, the $R_S$ rises, affecting the $FF$ and efficiency of the device [83,84].

### 3.2.8 Capacitance and Mott-Schottky analysis

**Figures 11(a)** and **(b)** demonstrate the variation in capacitance and Mott-Schottky (*MS*) values, respectively, when the voltage was enhanced from -0.5 V to 0.8 V for the six devices, and the frequency was fixed at 1 MHz. The capacitance progressively rose with the applied voltage and quickly improved at the higher voltages before reaching its maximum value. **Figure 11(a)** illustrates that all the device was depleted at zero bias, whereas when a forward bias of about 0.5 V was introduced, the depletion width declined to a value that was roughly equivalent to the absorber thickness. Therefore, as the forward bias voltage was increased, the capacitance improved as well and behaved following the *MS* relationship. It has previously been observed that at low voltages, the current is drastically lower than the saturation current, but at voltage spikes, the current is only allowed to reach the contact's saturation current [85]. On the other hand, the *MS* is a widely used and reliable tool for determining a device's built-in potential ($V_{bi}$), which is the contrast observed between the functions of electrode operation and doping level. The *MS* theory is primarily based on the characteristics of the p-n junction, where the x-axis intercept typically represents the $V_{bi}$ of the semiconductor devices. The slope of 1/C$^2$ (V) interprets as the concentration of inhabited trapping centers, even though the results are lower than anticipated due to the different electrode work functions [86]. For every device in this work, with the increase of applied voltage values, the *MS* values got lowered, which is similar to the previous studies [86,87].

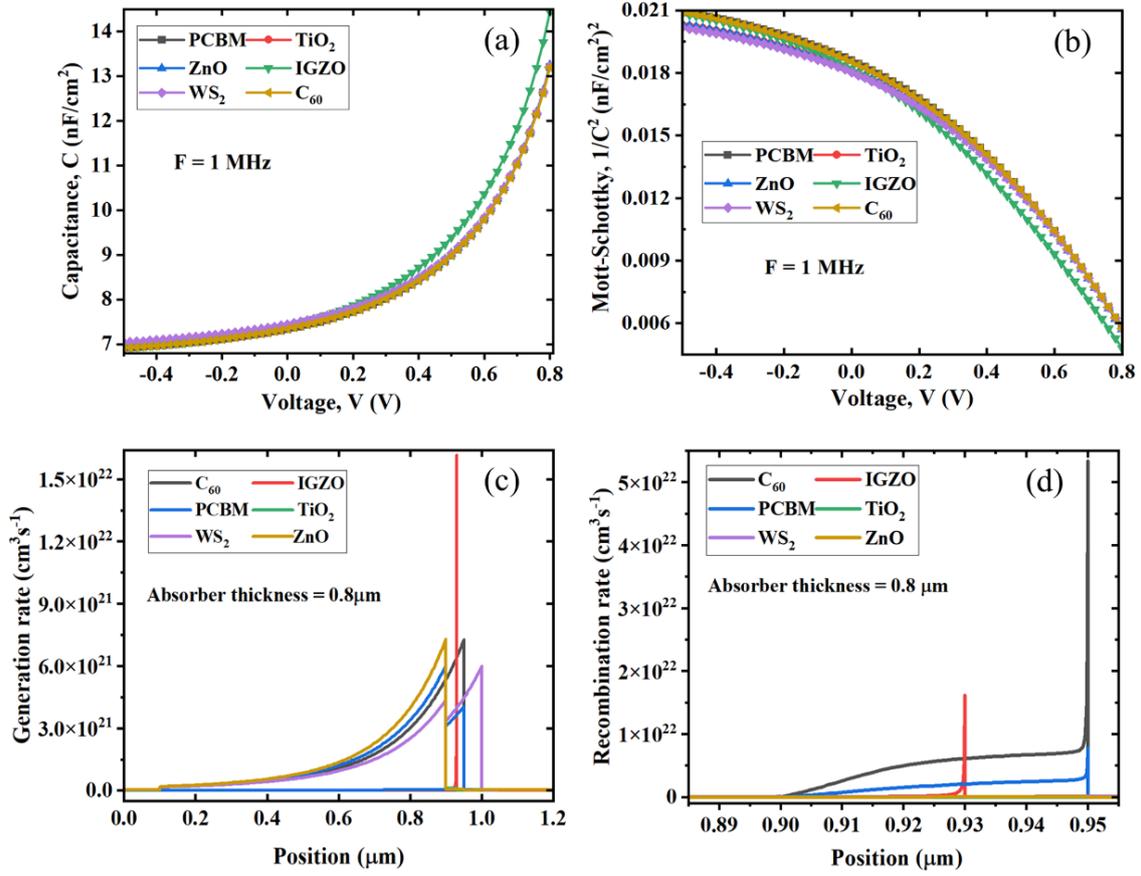

**Figure 11**: Analysis of (a) capacitance (C), (b) Mott-Schottky (1/C$^2$), (c) generation rate, and (d) recombination rate for CsPbI$_3$-based PSCs with different ETLs.

*3.2.9    Generation and recombination rate analysis*

**Figures 11(c)** and **(d)**, respectively, demonstrate the carrier generation and recombination rate concerning the position of the CsPbI$_3$-based PSCs. For all PSCs, the electron-hole pairs are formed during the carrier generation stage when an electron is excited from the valence band to the conduction band, leaving a hole in the valence band. The carrier formation is driven by the release of electrons and holes of these PSCs. Usually, the maximum generation rates for the devices were found when the position was in the 0.9 μm to 1.0 μm range, while IGZO ETL showed the maximum generation rate. Because at this range, the generation rate corresponding to the maximum number of electrons generated at that particular position in the device due to the increased absorption of photons in comparison with other positions. The creation of electron-hole pairs, G(x) is computed by SCAPS-1D using the incoming photon flux, N$_{phot}$ (λ, x), and **Eq. (11)** illustrates the value of G(x) in terms of this photon flux for every spectrum and each region:

$$G(\lambda, x) = \alpha(\lambda, x) \cdot N_{phot}(\lambda, x) \qquad (11)$$

On the contrary, the recombination rate is the opposite of the generation rate, in which the electrons and holes of the conduction band get united as well as eliminated. The lifespan and density of the charge carrier affect the rate of recombination in the PSCs. Moreover, the defect states that occur within each layer of PSC also influence electron-hole recombination. Similar to the generation rate, the maximum recombination rates for the devices were observed when the position was in the 0.9 μm to 1.0 μm range, where C$_{60}$ ETL showed the maximum recombination rate. Because at this particular range, in comparison with other positions, more electrons in the conduction band surpassed the energy gap and reached the valence band to become stable and occupy the position of a hole in the valence band. The electron-hole recombination profile within the devices is subsequently impacted by the energy levels that are produced at that point, and the grain boundaries and imperfections enable the recombination rate distribution to be non-uniform in the PSCs [87].

*3.2.10    J-V and QE characteristics*

**Figure 12(a)** shows the *J-V* characteristics for the six best configurations of this work. It was observed that the $TiO_2$, ZnO, and $WS_2$ as the ETL showed better *J-V* characteristics for the PSC in comparison with other ETLs, while the $C_{60}$ ETL showed the worst *J-V* characteristics.

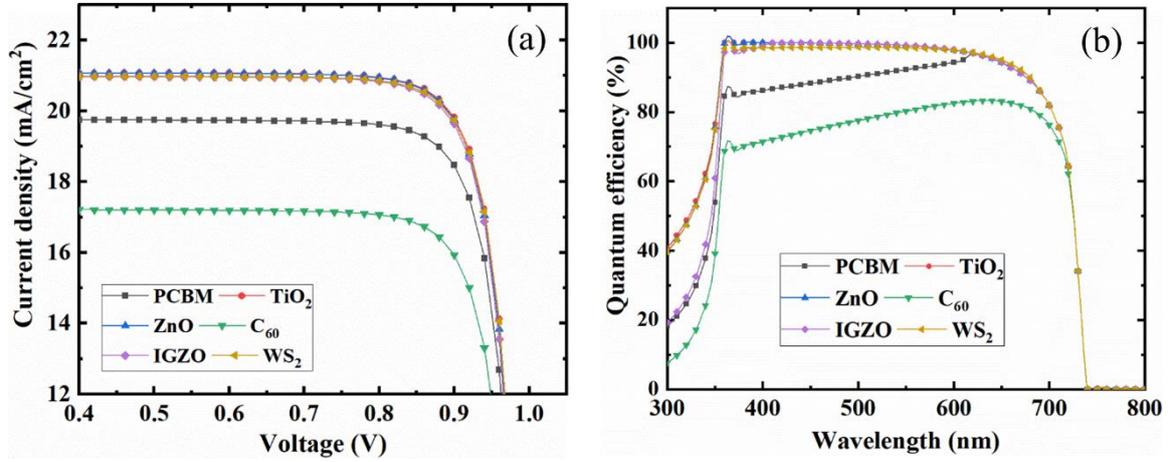

**Figure 12.** Effect of (a) *J-V*, and (b) *QE* for six studied devices.

**Figure 12(b)** illustrates how the wavelength and QE are interrelated for the six best devices of this study. The *QE* depends on the wavelength of the light (*λ*). It is the ratio of the number of charge carriers made by a solar cell to the number of photons that hit that cell [50]. The *QE* usually gets better as the absorber thickness gets higher as a thicker absorber can absorb more photons [82]. Similar to the *J-V* characteristics, when the $TiO_2$, ZnO, and $WS_2$ were utilized as the ETL, the PSC showed better *QE* characteristics, whereas the $C_{60}$ as the ETL showed the poorest *QE* characteristics.

**3.3    Comparison with wxAMPS results and previous work**

*3.3.1    Comparison between SCAPS-1D and wxAMPS results*

The findings of the SCAPS-1D (**Table 4**) were also validated through the simulations in wxAMPS (version 2.0) software. Both software conducted simulations with an absorber thickness, absorber acceptor density, and defect density of 800 nm, $10^{15}$ cm$^{-3}$, and $10^{15}$ cm$^{-3}$, respectively, to identify the change in PV characteristics among all 6 optimized devices. Similar to the previous SCAPS-1D simulations of this work, all the device simulations with wxAMPS were also run at an operating temperature of 300 K with an AM1.5G sun spectrum. A comparison between the SCAPS-1D and wxAMPS software simulation results has been enlisted in **Table 5**. The difference in results between the two simulation tools was satisfactory, and these kinds of findings were quite similar to previous studies [47,48,88].

**Table 5**: Comparison between SCAPS-1D and wxAMPS software simulation results.

| Device structure | Software | $V_{oc}$ (V) | $J_{sc}$ (mA/cm$^2$) | FF (%) | PCE (%) |
|---|---|---|---|---|---|
| ITO/PCBM/CsPbI$_3$/CBTS/Au | SCAPS-1D | 0.994 | 19.77 | 85.04 | 16.71 |
|  | wxAMPS | 1.000 | 21.00 | 84.51 | 17.77 |
| ITO/TiO$_2$/CsPbI$_3$/CBTS/Au | SCAPS-1D | 0.997 | 21.07 | 85.21 | 17.90 |
|  | wxAMPS | 1.010 | 22.97 | 85.23 | 19.75 |
| ITO/ZnO/CsPbI$_3$/CBTS/Au | SCAPS-1D | 0.997 | 21.07 | 85.03 | 17.86 |
|  | wxAMPS | 1.000 | 21.94 | 84.66 | 18.63 |
| ITO/C$_{60}$/CsPbI$_3$/CBTS/Au | SCAPS-1D | 0.989 | 17.25 | 84.80 | 14.47 |
|  | wxAMPS | 1.000 | 21.22 | 84.54 | 17.97 |
| ITO/IGZO/CsPbI$_3$/CBTS/Au | SCAPS-1D | 0.995 | 20.98 | 85.13 | 17.76 |
|  | wxAMPS | 1.000 | 22.82 | 84.65 | 19.36 |
| ITO/WS$_2$/CsPbI$_3$/CBTS/Au | SCAPS-1D | 0.997 | 20.98 | 85.22 | 17.82 |
|  | wxAMPS | 1.000 | 21.05 | 84.55 | 17.82 |

### 3.3.2 Comparison of SCAPS-1D results with previous work

**Table 6** shows the comparison between the recent experimental and theoretical works on $CsPbI_3$-based PSC with our study. All the previous experimental works have been done by the change of strategies like ionic incorporation, metastable phase, reducing crystal size/dimensional engineering, or steric hindrance to overcome the drawbacks of $CsPbI_3$ [22,89,98–102,90–97]. In those experiments, the PCE reached up to 18.40% [90] with these modifications. However, there is a lack of theoretical study available regarding the performance enhancement of the $CsPbI_3$ absorber, and the maximum PCE obtained via simulation is 15.6% [24]. We also noticed that most experiments used $TiO_2$, PTAA, and ZnO as the ETL, but there were variations in HTL and back contact metal. But in many cases, the results were not up to expectations. That's why we had gone through all these simulations to find out the best combinations to achieve an excellent performance.

**Table 6.** The comparison of PV parameters of $CsPbI_3$-based solar cells.

| Type | Device structure | Voc (V) | Jsc (mA/cm$^2$) | FF (%) | PCE (%) | Year | Ref. |
|---|---|---|---|---|---|---|---|
| E | FTO/TiO$_2$/CsPbI$_3$/PTAA/Au | 1.084 | 19.72 | 75.70 | 16.07 | 2019 | 89 |
| E | FTO/TiO$_2$/CsPbI$_3$/PTAA/Au | 1.090 | 20.34 | 77.00 | 17.03 | 2017 | 91 |
| E | FTO/TiO$_2$/CsPbI$_3$/P3HT/Au | 1.040 | 16.53 | 65.70 | 11.30 | 2018 | 92 |
| E | FTO/TiO$_2$/CsPbI$_3$/PTAA/Au | 1.059 | 18.95 | 75.10 | 15.07 | 2018 | 22 |
| E | FTO/TiO$_2$/CsPbI$_3$/C | 0.790 | 18.50 | 65.00 | 9.50 | 2018 | 93 |
| E | ITO/PTAA/CsPbI$_3$/C$_{60}$/BCP/Cu | 0.960 | 17.50 | 73.00 | 12.50 | 2019 | 94 |
| E | FTO/TiO$_2$/ CsPbI$_3$/P3HT/Au | 1.020 | 17.40 | 79.40 | 14.10 | 2018 | 95 |
| E | ITO/PTAA/CsPbI$_3$/BCP/Cu | 1.120 | 17.10 | 70.01 | 13.40 | 2019 | 96 |
| E | FTO/TiO$_2$/CsPbI$_3$/Spiro/Ag | 1.110 | 20.23 | 82.00 | 18.40 | 2019 | 90 |
| E | FTO/TiO$_2$/CsPbI$_3$/Spiro/MoO$_x$/Al | 0.990 | 13.47 | 65.00 | 10.77 | 2016 | 97 |
| E | FTO/TiO$_2$/CsPbI$_3$/Spiro/Au | 1.110 | 14.88 | 65.00 | 10.74 | 2018 | 98 |
| E | FTO/TiO$_2$/CsPbI$_3$/Spiro/Ag | 0.660 | 11.92 | 52.47 | 4.13 | 2016 | 99 |
| E | FTO/TiO$_2$/CsPbI$_3$/PTAA/Au | 0.993 | 19.51 | 70.46 | 13.65 | 2019 | 100 |
| E | FTO/TiO$_2$/CsPbI$_3$/Spiro/Au | 1.110 | 14.80 | 74.00 | 12.15 | 2019 | 102 |
| E | FTO/TiO$_2$/CsPbI$_3$/PTAA/Au | 1.204 | 15.25 | 78.70 | 14.45 | 2018 | 101 |
| T | FTO/ZnO/CsPbI$_3$/CuSbS$_2$/Se | 1.103 | 22.59 | 62.64 | 15.60 | 2021 | 24 |
| T | ITO/PCBM/CsPbI$_3$/CBTS/Au | 0.994 | 19.77 | 85.04 | 16.71 | 2022 | * |
| T | ITO/TiO$_2$/CsPbI$_3$/CBTS/Au | 0.997 | 21.07 | 85.21 | 17.90 | 2022 | * |
| T | ITO/ZnO/CsPbI$_3$/CBTS/Au | 0.997 | 21.07 | 85.03 | 17.86 | 2022 | * |
| T | ITO/C$_{60}$/CsPbI$_3$/CBTS/Au | 0.989 | 17.25 | 84.80 | 14.47 | 2022 | * |
| T | ITO/IGZO/CsPbI$_3$/CBTS/Au | 0.995 | 20.98 | 85.13 | 17.76 | 2022 | * |
| T | ITO/WS$_2$/CsPbI$_3$/CBTS/Au | 0.997 | 20.98 | 85.22 | 17.82 | 2022 | * |

Note: E = Experimental, T = Theoretical, *This work

## 4 Conclusion

In this study, first-principle DFT simulations were done to examine the structural and optical characteristics, as well as electronic properties of the $CsPbI_3$ absorber, including electron charge density, Fermi surface, DOS, and band structure. The $CsPbI_3$ absorber's computed bandgap using DFT was 1.483 eV. Additionally, both the charge density map and DOS demonstrated that the Pb-5d orbital is the primary source of contribution for the Pb atom. The observation of a surface that resembles a Fermi surface for holes and electrons indicates that the $CsPbI_3$ perovskite is multiband in nature. Furthermore, SCAPS-1D was used to perform numerical simulation and optimization of $CsPbI_3$-based PSC. A diverse set of HTLs and ETLs were explored to find the optimal combination from 96 device structures. From the optimization, CBTS HTL exhibited superior performance with PCBM, TiO$_2$, ZnO, C$_{60}$, IGZO, and WS$_2$ ETLs in comparison to other HTLs that were studied. We elected the six best devices to find the effect of absorber and ETL thickness, $R_S$, $R_{SH}$, and temperature on their performance along with their corresponding capacitance, Mott-Schottky, generation rate, recombination rate, $J$-$V$, and $QE$ characteristics. Among all six devices, the PSC with $TiO_2$ ETL and CBTS HTL showed the highest PCE of 17.9%. Moreover, the $TiO_2$ and ZnO ETLs showed the best performance during the absorber and ETL thickness variation. Due to the increment of $R_S$ and temperature, the performance of all six devices decreased, while the performance of each device increased with the increasing $R_{SH}$. The $C$-$V$ characteristics of the device with IGZO ETL were slightly different compared to other devices, whereas among all the devices, the generation and recombination rates of the device with IGZO ETL were better. When the $TiO_2$, ZnO, and $WS_2$ were utilized as the ETL, the PSC showed better $QE$ and $J$-$V$ characteristics, whereas the $C_{60}$ as the ETL showed the worst $QE$ and $J$-$V$ characteristics. Also, wxAMPS numerical simulations were carried out to compare with the SCAPS-1D simulation results. Finally, this simulation study assisted in finding an optimum ETL/$CsPbI_3$/HTL combination as a guide to experimentally fabricate a low-cost, stable, and highly efficient $CsPbI_3$-based PSC because it is impractical to experimentally investigate all probable combinations. The incorporation of machine learning into this work can be recommended to find a better combination for $CsPbI_3$-based PSC in the future.

**Supporting Information**

Energy level alignment of studied ITO, 9 ETLs, $CsPbI_3$, and 12 HTLs (Figure S1); Best optimized structures of $CsPbI_3$-based PSC with CBTS as the HTL and PCBM, $TiO_2$, ZnO, $C_{60}$, IGZO, and $WS_2$ as ETLs (Figure S2); Effect of the variation in temperature from 275 K to 475 K on $V_{OC}$, $J_{SC}$, $FF$, and PCE (Figure S3).

**Data availability**

The raw/processed data required to reproduce these findings cannot be shared at this time as the data also forms part of an ongoing study.

**Declaration of interests**

The authors declare that they have no known competing financial interests or personal relationships that could have appeared to influence the work reported in this paper.

**Funding Sources**

This research did not receive any specific grant from funding agencies in the public, commercial, or not-for-profit sectors.

**Acknowledgments**

The SCAPS-1D program was kindly provided by Dr. M. Burgelman of the University of Gent in Belgium. The authors would like to express their gratitude to him. They would also like to thank Professor A. Rockett and Dr. Yiming Liu from UIUC, as well as Professor Fonash from Penn State University, for their contributions to the wxAMPS program.

# SUPPORTING INFORMATION

**Table of content**

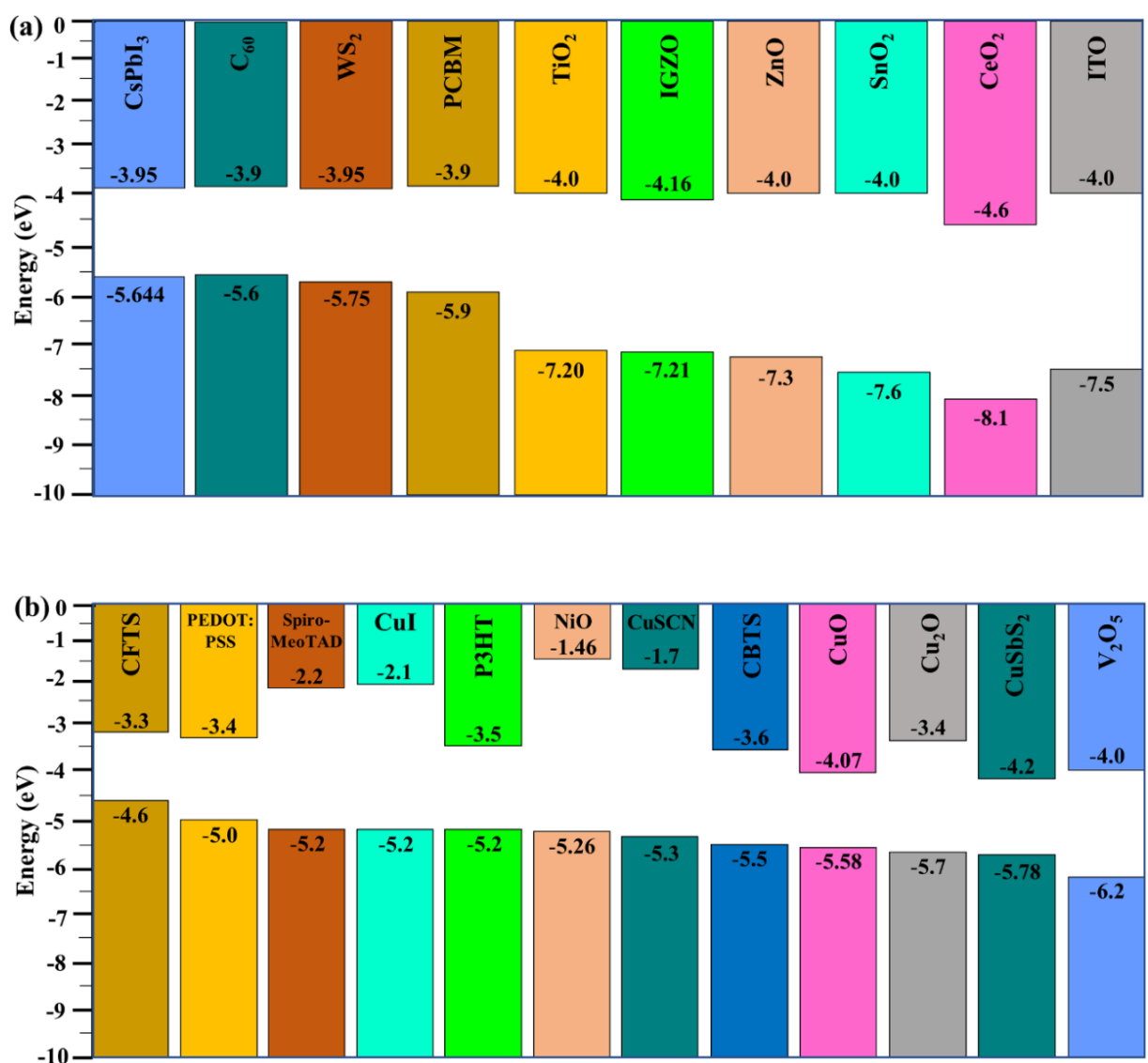

**Figure S1**: Energy level alignment of studied (a) ITO, ETLs, and absorber CsPbI$_3$, and (b) HTLs.

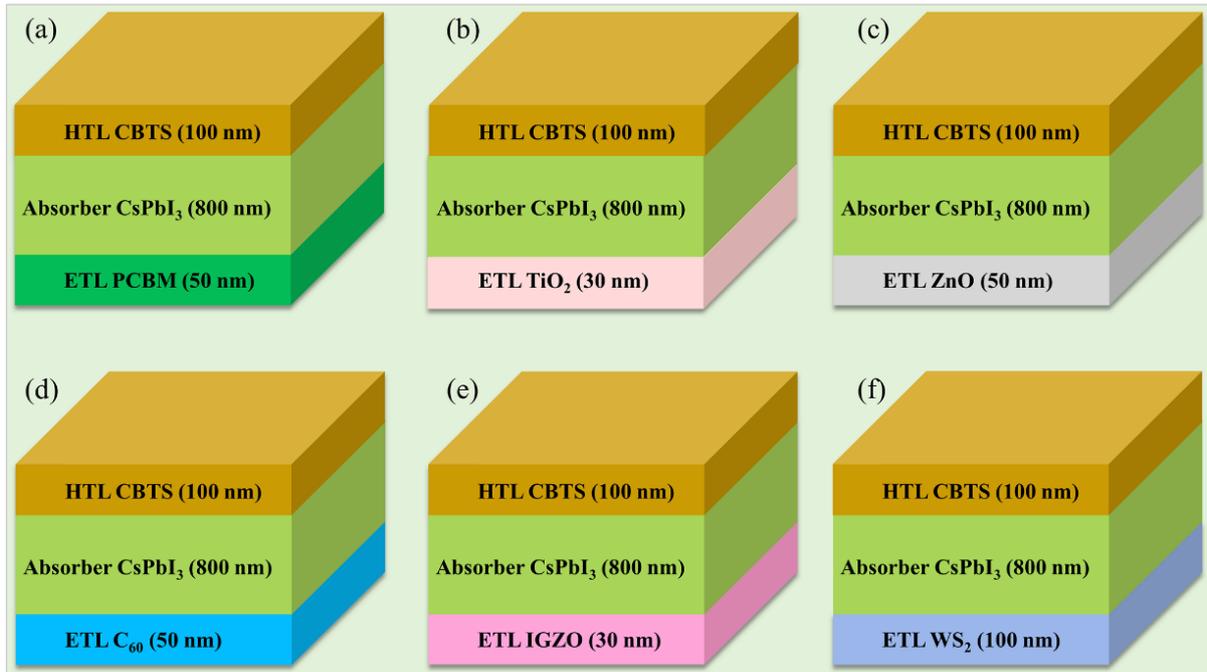

**Figure S2.** Best optimized structures of CsPbI$_3$-based PSC with CBTS as the HTL and (a) PCBM, (b) TiO$_2$, (c) ZnO, (d) C$_{60}$, (e) IGZO, and (f) WS$_2$ as the ETLs.

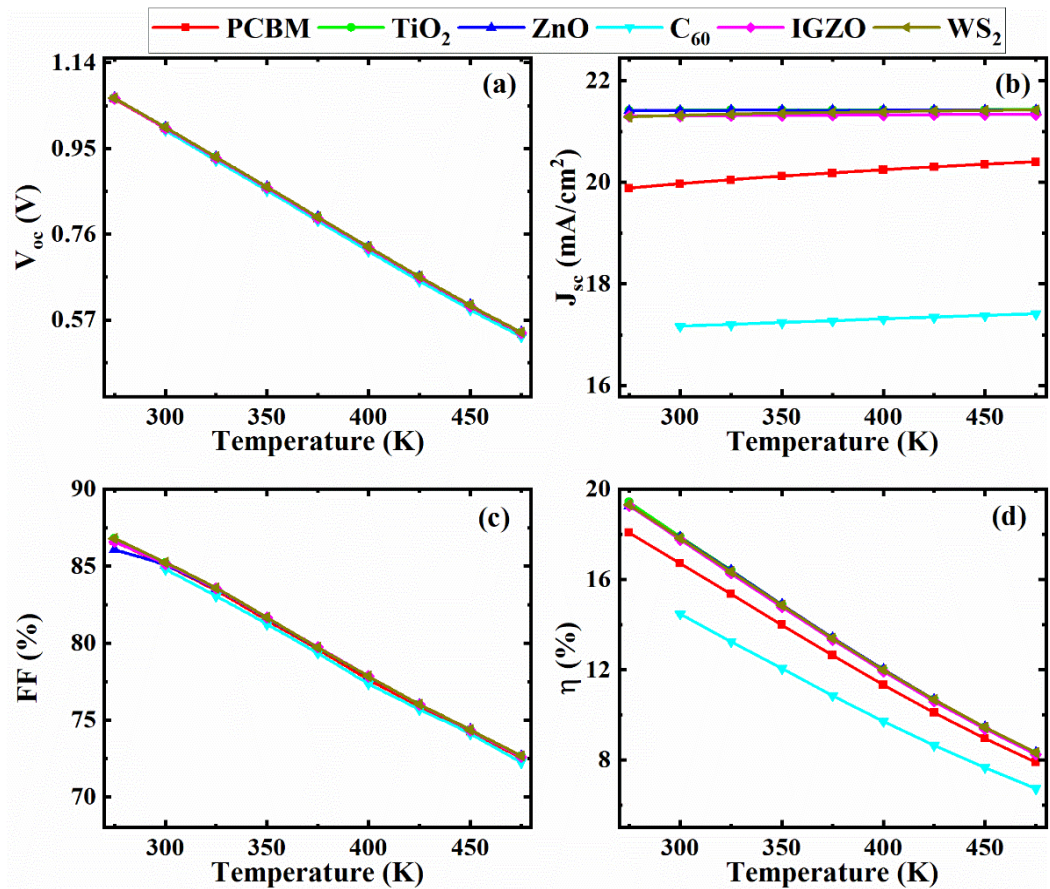

**Figure S3.** Effect of the variation in temperature from 275 K to 475 K on (a) $V_{OC}$, (b) $J_{SC}$, (c) *FF*, and (d) PCE.